\documentclass[12pt] {article}
\usepackage{psfrag}
\usepackage{graphicx}
\usepackage{latexsym,amsfonts}
\usepackage{amssymb}
\begin{document}
\setcounter{page}{1}
\def\theequation{\arabic{section}.\arabic{equation}}
\def\theequation{\thesection.\arabic{equation}}
\setcounter{section}{0}

\title{On strangeness production in  the reactions\\ $pp \to p
  \Lambda^0K^+$ and $pn \to n\Lambda^0K^+$ \\ near  threshold}

\author{A. N. Ivanov$^{1,2,3}$\,\thanks{Corresponding author. E--mail:
    ivanov@kph.tuwien.ac.at, Tel.: +43--1--58801--14261, Fax:
    +43--1--58801--14299}~\thanks{Permanent Address: State Polytechnic
    University, Department of Nuclear Physics, 195251 St. Petersburg,
    Russian Federation}, Ya. A. Berdnikov$^3$, M. Faber$^1$,\\
  V. A. Ivanova$^3$, J.
  Marton$^2$\,\thanks{http://www.oeaw.ac.at/smi}, N. I.
  Troitskaya$^3$}

\date{\today}

\maketitle

\begin{center}
{\it  $^1$Atominstitut der \"Osterreichischen Universit\"aten, Technische
Universit\"at Wien,  Wiedner Hauptstrasse 8-10, A-1040 Wien,
\"Osterreich \\  $^2$Stefan Meyer Institut f\"ur subatomare Physik, 
\"Osterreichische Akademie der Wissenschaften,
Boltzmanngasse 3, A-1090, Wien, \"Osterreich\\  $^3$State Polytechnic 
University, Polytechnicheskaya 29,\\
195251, St Petersburg, Russian Federation} 
\end{center}

\begin{center}
\begin{abstract}
  The cross sections for the reactions $pp \to p\Lambda^0K^+$ and $pn
  \to n\Lambda^0K^+$ are calculated near threshold of the final
  states.  The theoretical ratio of the cross sections $R = \sigma(pn
  \to n\Lambda^0K^+)/\sigma(pp \to p\Lambda^0K^+) \simeq 3$ shows the
  enhancement of the $pn$ interaction with respect to the $pp$
  interaction near threshold of the strangeness production
  $N\Lambda^0K^+$. Such an enhancement is caused by the contribution
  of the $np$ interaction in the isospin--singlet state, which is
  stronger than the $pn$ interaction in the isospin--triplet state.
  For the confirmation of this result we calculate the cross sections
  for the reactions $pp \to pp\pi^0$, $\pi^0 p \to \Lambda^0 K^+$ and
  $\pi^-p \to \Lambda^0 K^0$ near threshold of the final states. The
  theoretical cross sections agree well with the experimental data.
\end{abstract}
\end{center} 


\newpage

\section{Introduction}
\setcounter{equation}{0}

Recent experiments on the production of $K^+$\,--\, mesons in the
reaction $p d \to K^+ X N$, where $X$ is a hadronic state with
strangeness $S = - 1$ and baryon number $B = 2$ and $N$ is a nucleon,
have led to the prediction for the ratio of the cross sections for the
reactions $pp \to K^+ X $ and $pn \to K^+ X$
\begin{eqnarray}\label{label1.1}
  R = \frac{\sigma(pn \to K^+ X)}{\sigma(pp \to K^+ X)} \sim 3 - 4.
\end{eqnarray}
The estimate has been obtained for proton beam energies $T_p =
1.83\,{\rm GeV}$ and $T_p = 2.02\,{\rm GeV}$ giving $R \sim 3$ and $R
\sim 4$, respectively, to fit the experimental data on the
differential cross section for the reaction $p d \to K^+ X N$
\cite{COSY}. The parameter $R$ has been used as an input parameter.
This has been justified by the fact that the scattering of the proton
by the deuteron can be treated in the {\it impulse} approximation
\cite{TE88}, since there is no indication for any collective target
behaviour \cite{COSY} (see also \cite{EC94}).

Since the main contribution to the cross sections for the reactions
$pp \to K^+ X$ and $pn \to K^+ X$ comes from the channels $pp \to
p\Lambda^0 K^+$ and $pn \to n \Lambda^0 K^+$ \cite{COSY,KT99}, the
ratio Eq.(\ref{label1.1}) can be rewritten as follows
\begin{eqnarray}\label{label1.2}
 R = \frac{\sigma(pn \to n \Lambda^0 K^+)}{\sigma(pp \to p \Lambda^0 K^+)} 
\sim 3 - 4.
\end{eqnarray}
Can such a ratio be valid near threshold of the final state
$N\Lambda^0K^+$ ? \cite{VK05}.

A theoretical explanation of the ratio $R \sim 3 - 4$ near threshold
of the final $N\Lambda^0 K^+$ state has been proposed in \cite{KT99}
and \cite{CW97}.  As has been pointed out in \cite{CW97}, the value $R
\sim 3$ can be mainly due to the dominant contribution of the
$\rho$\,--\,meson exchange in addition to the contribution of the
resonances $N(1535)$ and $N(1650)$ with the quantum numbers $I(J^P) =
\frac{1}{2}(\frac{1}{2}^-)$ \cite{PDG04}, described within the
Breit--Wigner approach \cite{CW05}.  In \cite{KT99} the cross sections
for the reactions $pN \to N\Lambda^0K^+$ have been calculated within
the resonance model of $NN$ scattering for energies of the incident
proton far from threshold.  Within such an approach the ratio $R$ can
be obtained assuming the validity of the cross sections in the
vicinity of threshold \cite{AS05}.  Recently \cite{AI01} the
strangeness production in $pp$ collisions $pp \to p\Lambda^0K^+$ and
$pp\to p\Sigma^0K^+$ near threshold of the final states has been
analysed by using chiral Lagrangians with linear realization of chiral
symmetry and non--derivative meson--baryon couplings
\cite{HP73}--\cite{GG69}. The matrix elements of the transitions $pp
\to p\Lambda^0K^+$ and $pp \to p\Sigma^0 K^+$ have been calculated in
the one--meson exchange approximation \cite{NK99}. For the calculation
of the amplitudes of the reactions $pp \to p\Lambda^0K^+$ and $pp \to
p\Sigma^0 K^+$ there has been taken into account the $pp$ rescattering
in the initial state, i.e. the initial state interaction \cite{KN99},
and $p\Lambda^0K^+$ and $p\Sigma^0K^+$ final--state interaction
\cite{FSI,FSIa}. As has been shown in \cite{IV4} the same analysis of
the strangeness production has turned out to be very useful for the
correct description of the reactions $K^-d \to NY$ of $K^-d$
scattering at threshold, where $NY = p\Sigma^-$, $n\Sigma^0$ and
$n\Lambda^0$. In this paper we apply such an analysis of the
strangeness production to the calculation of the cross sections for
the reactions $pN \to N\Lambda^0K^+$ near threshold of the final state
$N\Lambda^0K^+$ and the ratio $R$.

The paper is organised as follows. In Section 2 we calculate the cross
section for the reaction $pp \to p\Lambda^0K^+$. The obtained result
agrees with the experimental data \cite{JB98} with an accuracy better
$16\,\%$. We show that the contribution of the vector $\rho$\,-- and
$\omega^0$\,--\,meson exchanges to the cross section for the reaction
$pp \to p\Lambda^0K^+$ makes up about $9\,\%$ only. In Section 3 we
calculate the cross section for the reaction $pn \to n\Lambda^0K^+$.
We calculate the cross sections for the reactions with the $pn$ pair
in the state with isospin $I = 1$ and isospin $I = 0$ and demonstrate
that $R^{(I = 1)} = \sigma((pn)_{I = 1} \to n\Lambda^0K^+)/\sigma(pp
\to p\Lambda^0K^+) \simeq 1$.  We find that the contribution of the
vector $\rho$\,-- and $\omega^0$\,--\,meson exchanges to the cross
section for the reaction $(pn)_{I = 1} \to n\Lambda^0K^+$ with the
$pn$ pair in the isospin--triplet state is about $3\,\%$.  In turn,
the vector\,--\,meson exchanges do not influence on the value of the
cross section for the reaction $(pn)_{I = 0} \to n\Lambda^0K^+$ with
the $pn$ pair in the isospin--singlet state.  In Section 4 we
calculate the ratio $R$ of the cross sections for the reactions $pn
\to n\Lambda^0K^+$ and $pp \to p\Lambda^0K^+$.  We get $R \simeq 3$.
This agrees with the enhancement of the $pn$ interaction with respect
to the $pp$ interaction in the reactions of the strangeness production
observed by the ANKE--Collaboration at COSY and the theoretical
prediction in \cite{KT99} and \cite{CW97}. We show that such an
enhancement is caused by the contribution of $pn$ interaction in the
isospin--singlet state $(pn)_{I = 0}$, which is stronger than the $pn$
interaction in the isospin--triplet state $(pn)_{I = 1}$. In Section 5
we analyse the contribution of the resonances $N(1535)$ and $N(1650)$,
treating them as elementary particles \cite{MR96}--\cite{IV9}.  We
show that the contribution of these resonances to the amplitudes of
the reactions $pN \to N\Lambda^0K^+$ can be neglected with respect to
the contribution of the ground state baryon octet coupled to the
octets of the pseudoscalar and scalar mesons. In Section 6 we
calculate the cross section for the reaction $pp \to pp\pi^0$, caused
by the pure ${^3}{\rm P}_0 \to {^1}{\rm S}_0$ transition with
pseudoscalar and scalar meson exchanges. Near threshold the
theoretical cross section agrees well with the experimental data.  In
Section 7 we calculate the cross sections for the reactions $\pi^0p\to
\Lambda^0 K^+$ and $\pi^-p \to \Lambda^0 K^0$. We show that the
theoretical cross section for the reaction $\pi^-p \to \Lambda^0 K^0$
agrees well with the experimental data for the energy excess
$\varepsilon \le 13\,{\rm MeV}$ \cite{JJ71} and the results obtained
within $SU(3)$ chiral dynamics with coupled channels \cite{WW97}.  We
show that for the correct description of the cross sections for the
reactions $\pi^0p\to \Lambda^0 K^+$ and $\pi^-p \to \Lambda^0 K^0$ the
contribution of the resonances $N(1535)$ and $N(1650)$ as well as
$\Sigma(1750)$ is very important due to the cancellation of the
non--resonance contribution. The dominant role of resonances $N(1535)$
and $N(1650)$ in the amplitudes of the reactions $\pi^0p\to \Lambda^0
K^+$ and $\pi^-p \to \Lambda^0 K^0$ has been pointed out by Kaiser
{\it et al.} \cite{WW97}. In the Conclusion we discuss the obtained
results.  In Appendix A we analyse the contribution of the
final--state interaction to the amplitudes of the reactions $pN \to
N\Lambda^0K^+$.

\section{Cross section for the reaction $pp \to p \Lambda^0K^+$  near
  threshold}
\setcounter{equation}{0}

Following \cite{KT99,CW97} and \cite{AI01,NK99,IV4} we define the
amplitude of the transition $pp \to p\Lambda^0 K^+$ in the
tree--approximation with one--meson exchanges. A complete set of
Feynman diagrams, describing the transition $pp \to p\Lambda^0 K^+$
near threshold of the final $p\Lambda^0 K^+$ state, is depicted in
Fig.1 and Fig.2.
\begin{figure}
\centering 
\psfrag{K+}{$K^+$} 
 \psfrag{K-}{$K^-$} 
\psfrag{L0}{$\Lambda^0$} \psfrag{p0}{$\pi^0$}
\psfrag{S0}{$\Sigma^0$}
\psfrag{Sp}{$\Sigma^+$}\psfrag{p}{$p$} 
\psfrag{e}{$\eta$}
\psfrag{s}{$\sigma$}
\psfrag{d}{$\delta^0$}
\psfrag{k+}{$\kappa^+$}
\psfrag{k-}{$\kappa^-$}
\includegraphics[height= 0.90\textheight]{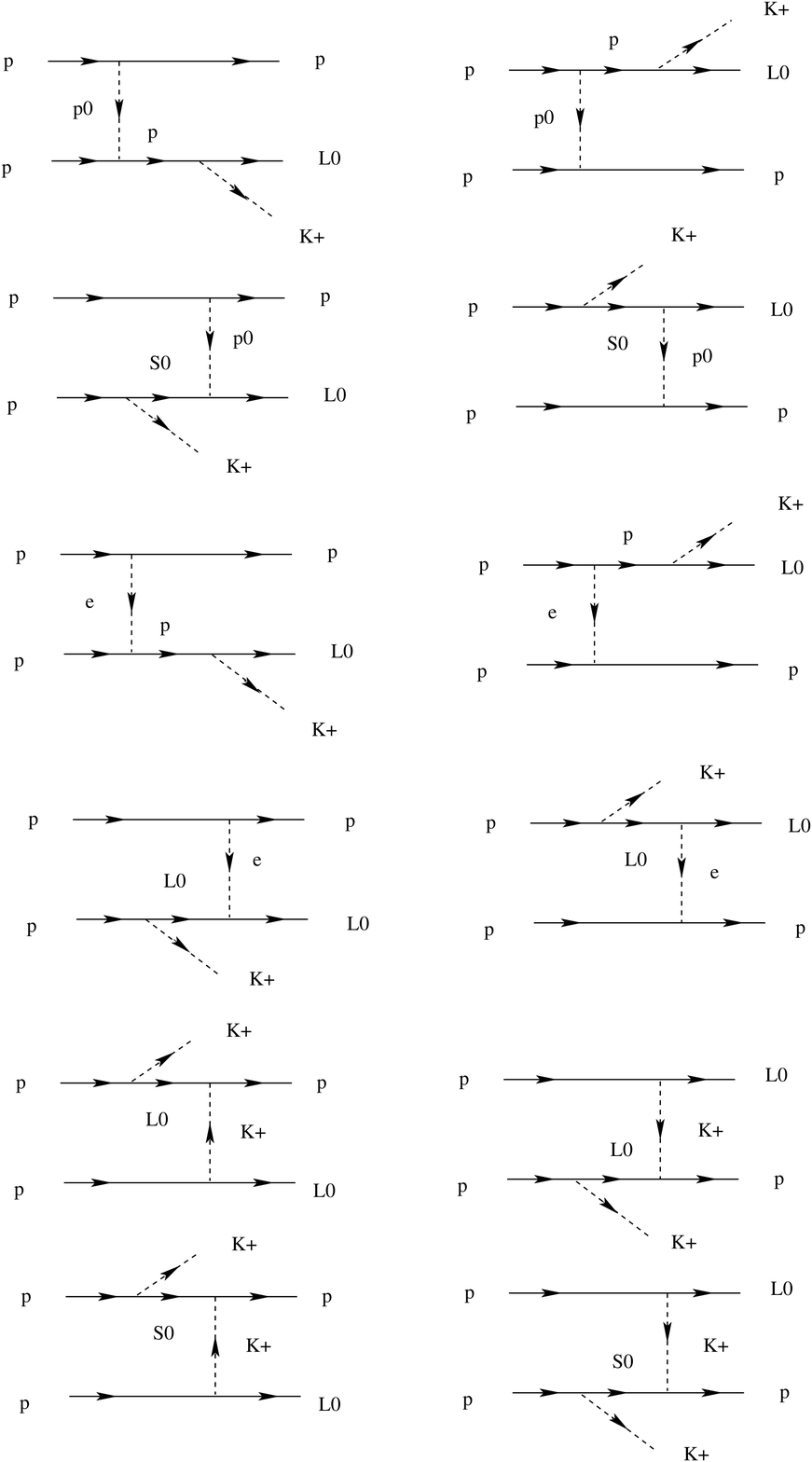}
\caption{The set of Feynman diagrams contributing to the effective
  coupling constant $C^{\,pp}_{p\Lambda^0K^+}$ of the $p p \to p
  \Lambda^0K^+$ transition in the one--pseudoscalar meson exchange
  approximation.}
\end{figure}
\begin{figure}
\centering 
\psfrag{K+}{$K^+$} 
 \psfrag{K-}{$K^-$} 
\psfrag{L0}{$\Lambda^0$} \psfrag{p0}{$\pi^0$}
\psfrag{S0}{$\Sigma^0$}
\psfrag{Sp}{$\Sigma^+$}\psfrag{p}{$p$} 
\psfrag{e}{$\eta$}
\psfrag{s}{$\sigma$}
\psfrag{d}{$\delta^0$}
\psfrag{k+}{$\kappa^+$}
\psfrag{k-}{$\kappa^-$}
\includegraphics[height= 0.60\textheight]{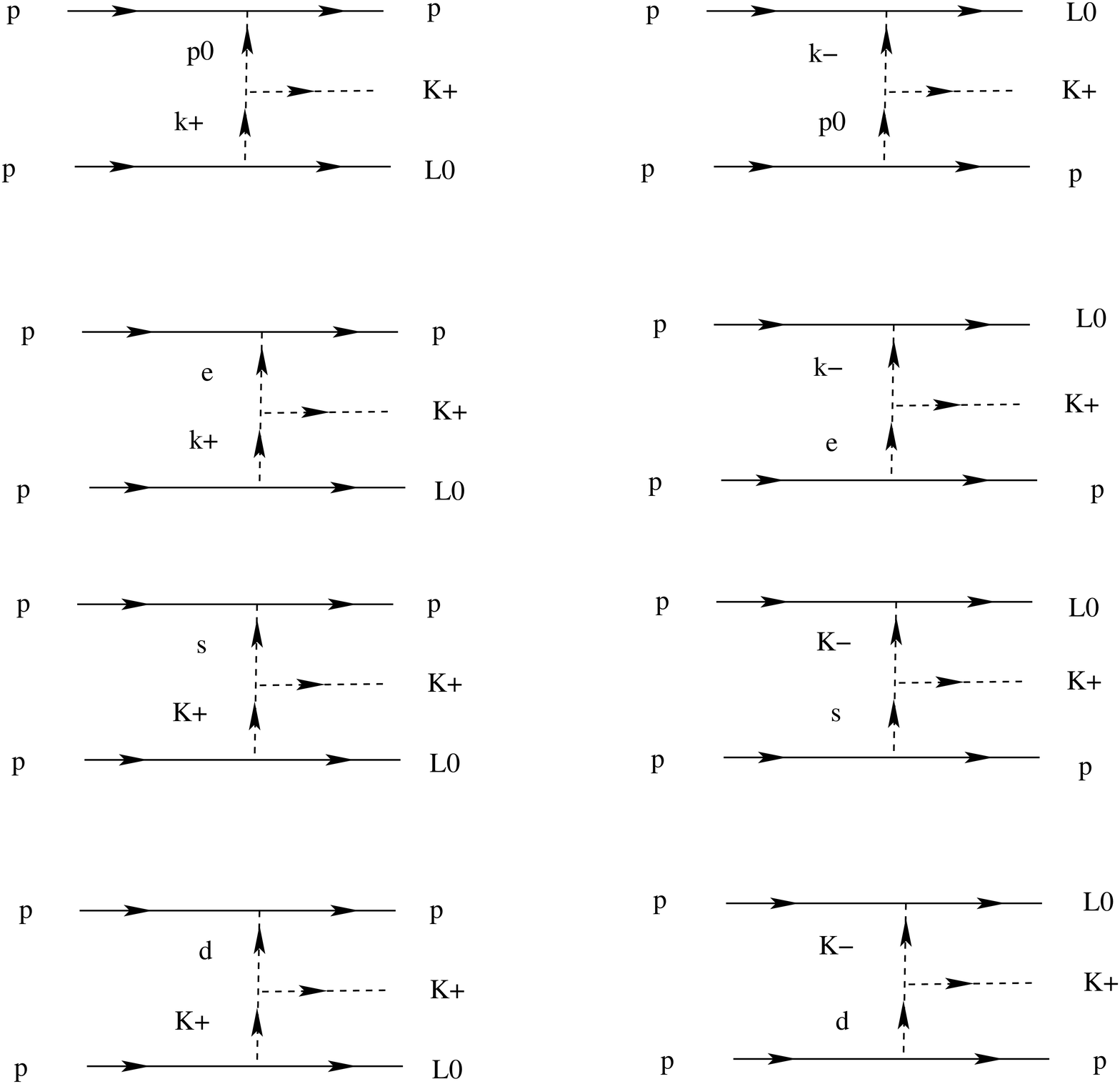}
\caption{The set of Feynman diagrams contributing to the effective
  coupling constant $C^{\,pp}_{p\Lambda^0K^+}$ of the $ p p \to p
  \Lambda^0 K^+$ transition in the one--scalar meson exchange
  approximation.}
\end{figure}
The calculation of these diagrams we carry out using chiral Lagrangian
with non--derivative meson--baryon couplings invariant under linear
transformations of chiral $SU(3)\times SU(3)$ symmetry
\cite{HP73}--\cite{GG69}.  Since masses of scalar partners of
pseudoscalar mesons are not well--defined, we calculate the amplitudes
in the limit of infinite masses of scalar mesons.  According to
\cite{SW67,BL69}, such a limit corresponds to the calculation of the
amplitudes with chiral Lagrangian invariant under non--linear
transformations of chiral $SU(3)\times SU(3)$ symmetry\cite{GG69}. Of
course, such an equivalence has been proved only in the
tree--approximation \cite{SW67}--\cite{GG69}.

For the derivation of the analytical expressions of these diagrams we
use the meson--baryon interactions (see Eqs.({\rm D}.4), ({\rm D}.7)
and ({\rm D}.9) of Ref.\cite{IV4} and also \cite{MN79}) and the wave
functions of the initial and the final states in the particle number
representation:
\begin{eqnarray}\label{label2.1}
  |p(\vec{p}_1,\sigma_1)p(\vec{p}_2,\sigma_2)\rangle &=& 
  a^{\dagger}_p(\vec{p}_1,\sigma_1)a^{\dagger}_p(\vec{p}_2,\sigma_2)|0\rangle,\nonumber\\
  |p(\vec{k}_p,\sigma_p)\Lambda^0(\vec{k}_{\Lambda},\sigma_{\Lambda})
K^+(\vec{Q}_K)\rangle &=& 
  a^{\dagger}_p(\vec{k}_p,\sigma_p)a^{\dagger}_{\Lambda}
  (\vec{k}_{\Lambda},\sigma_{\Lambda})c^{\dagger}_{K^+}(\vec{Q}_K)|0\rangle,
\end{eqnarray}
where creation (annihilation) operators of baryons and the
$K^-$\,--\,meson obey standard relativistic covariant anti--commutation
and commutation relations \cite{IV4}.

Near threshold of the final state $p\Lambda^0K^+$ the Feynman diagrams
in Fig.\,1 and Fig.\,2 can be described by the effective local
Lagrangian of the transition $pp \to p\Lambda^0K^+$ \cite{AI01,IV4}:
\begin{eqnarray}\label{label2.2}
{\cal L}^{pp \to p\Lambda^0K^+}_{\rm eff}(x) &=& A^{pp}_{p\Lambda^0K^+}
[\bar{\Lambda}^0(x)p(x)][\bar{p}(x)i\gamma^5 p(x)]\,K^{+\dagger}(x)\nonumber\\
&+& 
B^{pp}_{p\Lambda^0K^+}[\bar{\Lambda}^0(x)i\gamma^5 p(x)][\bar{p}(x)p(x)]\,K^{+\dagger}(x),
\end{eqnarray}
where the coefficients $A^{pp}_{p\Lambda^0K^+}$ and
$B^{pp}_{p\Lambda^0K^+}$ are equal to
\begin{eqnarray}\label{label2.3}
  A^{pp}_{p\Lambda^0K^+} &=&-\,\frac{1}{\sqrt{3}}\,\frac{(3 - 2\alpha_D)\,
    g^3_{\pi NN}}{m^2_{\pi} + 2m_N T_N}\,\frac{1}{m_N + m_{\Lambda^0} + m_K}
  \nonumber\\
  &&+\frac{2}{\sqrt{3}}\,\frac{\alpha_D\,(2\alpha_D - 1)\,g^3_{\pi NN}}{m^2_{\pi}
    + 2m_N T_N}\,\frac{m_{\Sigma} + m_K - m_N}{m^2_{\Sigma} + 2 m_K T_N 
    - (m_N - m_K)^2}
  \nonumber\\
  &&-\,\frac{1}{3\sqrt{3}}\,\frac{(3 - 2\alpha_D)(3- 4\alpha_D)^2\,g^3_{\pi NN}}{
    m^2_{\eta} + 2m_N T_N}\,\,\frac{1}{m_N + m_{\Lambda^0} + m_K}\nonumber\\
  &&+\frac{2}{3\sqrt{3}}\,\frac{\alpha_D\,(3 - 2\alpha_D)(3 - 4\alpha_D)\,
    g^3_{\pi NN}}{m^2_{\eta} + 2m_N T_N}\,\frac{m_{\Lambda^0} + m_K - m_N}{
    m^2_{\Lambda} + 2 m_K T_N - (m_N - m_K)^2}\nonumber\\
  && - \,\frac{1}{2\sqrt{3} g^2_A}\,\frac{(3 - 2\alpha_D)\,g^3_{\pi NN}}{m^2_{\pi} + 2 m_N T_N}\,
\frac{1}{m_N} + \,\frac{1}{6 g^2_A}\,\frac{(3 - 4\alpha_D)\,g^3_{\pi NN}}{m^2_{\eta} + 2 m_N T_N}\,
\frac{1}{m_N} = \nonumber\\
  &=&-\,1.93\times 10^{-6}\,{\rm MeV}^{-3},\nonumber\\
  B^{pp}_{p\Lambda^0K^+} &=&+\,\frac{1}{3\sqrt{3}}\,\frac{(3 - 2\alpha_D)^3
    \,g^3_{\pi NN}}{m^2_K + 2m_{\Lambda^0} T_N - (m_{\Lambda^0} - m_N)^2}\,
  \frac{m_{\Lambda^0} + m_K - m_N}{
    m^2_{\Lambda} + 2 m_K T_N - (m_N - m_K)^2}\nonumber\\
  &&+\,\frac{1}{\sqrt{3}}\,\frac{(2\alpha_D - 1)^2(3 - 2\alpha_D)\,
    g^3_{\pi NN}}{m^2_K +
    2m_{\Lambda^0} T_N - (m_{\Lambda^0} - m_N)^2}\,\frac{m_{\Sigma} 
    + m_K - m_N}{m^2_{\Sigma} + 2 m_K T_N - (m_N - m_K)^2}\nonumber\\
  &&-\,\frac{1}{\sqrt{3} g^2_A}\,\frac{(3 - 2\alpha_D)\,g^3_{\pi NN}}{m^2_K 
    + 2 m_{\Lambda^0} T_N - (m_{\Lambda^0} - m_N)^2}\,\frac{1}{m_N} =\nonumber\\
&=& -\,0.35\times 10^{-6}\,{\rm MeV}^{-3}, 
\end{eqnarray}
where $g_{\pi NN} = 13.21$ is the coupling constant of $\pi NN$
interactions \cite{PSI2}, $g_A = 1.267$ is the axial--vector coupling
constant, renormalized by strong low--energy interactions, $F_{\pi} =
92.4\,{\rm MeV}$ is the PCAC constant and $\alpha_D = D/(D + F) =
0.635$ is the Gell--Mann parameter \cite{PDG04}, $T_N = E_N - m_N =
(m_{\Lambda^0} + m_K - m_N)/2 = 335\,{\rm MeV}$ is the kinetic energy
of the relative motion of the protons at threshold in the center of
mass frame, calculated for $m_N = 940\,{\rm MeV}$, $m_{\Lambda^0} =
1116\,{\rm MeV}$ and $m_K = 494\,{\rm MeV}$.  We have used also
$m_{\Sigma} = 1193\,{\rm MeV}$, $m_{\pi} = 140\,{\rm MeV}$ and
$m_{\eta} = 550\,{\rm MeV}$ \cite{PDG04}.

By a Fierz transformation we get
\begin{eqnarray}\label{label2.4}
{[\bar{\Lambda}^0(x)p(x)][\bar{p}(x)i\gamma^5 p(x)]} &=& - \frac{1}{4}\,
[\bar{\Lambda}^0(x)i\gamma^5 p^c(x)][\bar{p^c}(x)p(x)]\nonumber\\
&& + \frac{1}{4}\,i\,
[\bar{\Lambda}^0(x)\gamma_{\mu} p^c(x)][\bar{p^c}(x)\gamma^{\mu} \gamma^5 p(x)]
 + \ldots,
\nonumber\\
{[\bar{\Lambda}^0(x)i\gamma^5p(x)][\bar{p}(x)p(x)]} &=& - \frac{1}{4}\,
[\bar{\Lambda}^0(x)i\gamma^5 p^c(x)][\bar{p^c}(x)p(x)]\nonumber\\
&& - \frac{1}{4}\,i\,
[\bar{\Lambda}^0(x)\gamma_{\mu} p^c(x)][\bar{p^c}(x)\gamma^{\mu} \gamma^5 p(x)]
 + \ldots\,.
\end{eqnarray} 
Hence, the effective Lagrangian of the transition $pp\to
p\Lambda^0K^+$ is equal to
\begin{eqnarray}\label{label2.5}
{\cal L}^{pp \to p\Lambda^0K^+}_{\rm eff}(x) &=& -\,\frac{1}{4}\,
C^{(pp)_{{^3}{\rm P}_0}}_{(p\Lambda^0)_{{^1}{\rm S}_0}K^+}\,
[\bar{\Lambda}^0(x)i\gamma^5 p^c(x)][\bar{p^c}(x)p(x)]\,K^{+\dagger}(x)\nonumber\\
&& + \,\frac{1}{4}\,
C^{(pp)_{{^3}{\rm P}_1}}_{(p\Lambda^0)_{{^3}{\rm S}_1}K^+}\,
[\bar{\Lambda}^0(x)i\gamma^{\mu} p^c(x)][\bar{p^c}(x)\gamma_{\mu}\gamma^5 p(x)]\,K^{+\dagger}(x),
\end{eqnarray}
where the first and the second terms describe the production of the
$p\Lambda^0$ pair in the ${^1}{\rm S}_0$ and ${^3}{\rm S}_1$ state by
the $pp$ pair in the ${^3}{\rm P}_0$ and ${^3}{\rm P}_1$ state,
respectively. The effective coupling constants $C^{(pp)_{{^3}{\rm
      P}_0}}_{(p\Lambda^0)_{{^1}{\rm S}_0}K^+}$ and $C^{(pp)_{{^3}{\rm
      P}_1}}_{(p\Lambda^0)_{{^3}{\rm S}_1}K^+}$ are equal to
\begin{eqnarray}\label{label2.6}
  C^{(pp)_{{^3}{\rm P}_0}}_{(p\Lambda^0)_{{^1}{\rm S}_0}K^+} &=&  
A^{pp}_{p\Lambda^0K^+} +  B^{pp}_{p\Lambda^0K^+} = -\,2.28\times 10^{-6}\,
{\rm MeV}^{-3},\nonumber\\
  C^{(pp)_{{^3}{\rm P}_1}}_{(p\Lambda^0)_{{^3}{\rm S}_1}K^+} &=& 
A^{pp}_{p\Lambda^0K^+} -  B^{pp}_{p\Lambda^0K^+} = -\,1.58\times 10^{-6}\,
{\rm MeV}^{-3}
\end{eqnarray}
The total cross section for the reaction $pp \to
p\Lambda^0K^+$ is given by
\begin{eqnarray}\label{label2.7}
  \hspace{-0.3in} &&\sigma^{pp\to p\Lambda^0K^+}(\varepsilon) = \sigma^{(pp)_{{^3}{\rm
        P}_0}\to (p\Lambda^0)_{{^1}{\rm S}_0}K^+}(\varepsilon) +
  \sigma^{(pp)_{{^3}{\rm P}_1}\to (p\Lambda^0)_{{^3}{\rm
        S}_1}K^+}(\varepsilon) =\nonumber\\
  \hspace{-0.3in} &&\hspace{1in} = \frac{1}{128\pi^2}\frac{\varepsilon^2}{m_K}\frac{p_0}{E_0}\,
  \Big(\frac{m_Km_{\Lambda^0}m_N}{m_K + m_{\Lambda^0} +
    m_N}\Big)^{3/2}\nonumber\\
  \hspace{-0.3in} &&
  \times \,\Big(\frac{1}{3}\,
  \Big|C^{(pp)_{{^3}{\rm P}_0}}_{(p\Lambda^0)_{{^1}{\rm S}_0}K^+}\Big|^2
  \Big|f_{(pp)_{{^3}{\rm
        P}_0}}(p_0)\Big|^2 + \frac{2}{3}\,
  \Big|C^{(pp)_{{^3}{\rm P}_1}}_{(p\Lambda^0)_{{^3}{\rm S}_1}K^+}\Big|^2
  \Big|f_{(pp)_{{^3}{\rm
        P}_1}}(p_0)\Big|^2\Big)\,\Omega_{p\Lambda^0 K^+}(\varepsilon),
\end{eqnarray}
where $\varepsilon = 2E_N - m_K - m_{\Lambda^0} - m_N$ is the excess
energy for the relative energy of the $pp$ pair in the center of mass
frame $E_N = \sqrt{p^{\;2} + m^2_N}$. It is measured in MeV.  The
function $\Omega_{p\Lambda^0 K^+}(\varepsilon)$ is related to the
account for the final--state interaction in the $p\Lambda^0$ and
$pK^+$ channels defined in Appendix A (see also \cite{AI01}). For
$\varepsilon \ge 6.68\,{\rm MeV}$ the contribution of the Coulomb
repulsion in the $K^+p$ pair makes up about $15\,\%$. Therefore for
$\varepsilon \ge 6.68\,{\rm MeV}$ we carry out the calculation of
$\Omega_{p\Lambda^0 K^+}(\varepsilon)$ at the neglect of the Coulomb
repulsion.

The amplitudes $f_{(pp)_{{^3}{\rm P}_0}}(p_0)$ and $f_{(pp)_{{^3}{\rm
      P}_1}}(p_0)$ describe the $pp$ rescattering in the ${^3}{\rm
  P}_0$ and ${^3}{\rm P}_1$ state at threshold of the final
$p\Lambda^0K^+$ state, respectively. The detailed procedure for the
calculation of the amplitudes $f_{(pp)_{{^3}{\rm P}_0}}(p_0)$ and
$f_{(pp)_{{^3}{\rm P}_1}}(p_0)$ is expounded in \cite{IV4}.  Following
this procedure one gets
\begin{eqnarray}\label{label2.8}
  |f_{(pp)_{{^3}{\rm P}_0}}(p_0)| &=& \Big|\Big\{1 + 
\frac{C_{(pp)_{{^3}{\rm P}_0}}(p_0)}{8\pi^2}\,
  \frac{p^3_0}{E_N}\,\Big[{\ell n}\Big(\frac{E_N + p_0}{E_N - p_0}\Big) 
  +\,i\,\pi\Big]\Big\}^{-1}\Big| = 0.15,\nonumber\\
  |f_{(pp)_{{^3}{\rm P}_1}}(p_0)| &=& \Big|\Big\{1 - 
  \frac{ C_{(pp)_{{^3}{\rm P}_1}}(p_0)}{12\pi^2}\,
  \frac{p^3_0}{E_N}\,\Big[{\ell n}\Big(\frac{E_N + p_0}{E_N - p_0}\Big)
  -\,i\,\pi\Big]\Big\}^{-1}\Big| = 0.24,
\end{eqnarray}
where the effective coupling constants $C_{(pp)_{{^3}{\rm P}_0}}(p_0)$
and $C_{(pp)_{{^3}{\rm P}_1}}(p_0)$ are defined by the $\pi^0$\,-- and
$\eta$\,--\,meson exchanges\,\footnote{By using our expression for the
  amplitude $f_{(pp)_{{^3}{\rm P}_0}}(p_0)$ and the results obtained
  in \cite{KN99} we can estimate the values of the phase shift
  $\delta_{{^3}{\rm P}_0}(p_0)$ and the inelasticity $\eta_{{^3}{\rm
      P}_0}(p_0)$ of $pp$ scattering in the ${^3}{\rm P}_0$ state. We
  get $\delta_{{^3}{\rm P}_0}(p_0) = -\,63.0^0$ and $\eta_{{^3}{\rm
      P}_0}(p_0) = 0.80$. This does not contradict the SAID analysis
  of the experimental data \cite{SAID1,SAID2}: $\delta_{{^3}{\rm
      P}_0}(p_0) = -\,62.7^0$ and $\eta_{{^3}{\rm P}_0}(p_0) =
  0.65$.}.

For the relative momentum of the $pp$ pair $p_0 = \sqrt{T_N(T_N +
  2m_N)} = 861\,{\rm MeV}$ at threshold of the reaction $pp \to
p\Lambda^0K^+$, the effective coupling constants $C_{(pp)_{{^3}{\rm
      P}_0}}(p_0)$ and $C_{(pp)_{{^3}{\rm P}_1}}(p_0)$ amount to
\cite{IV4}
\begin{eqnarray}\label{label2.9}
  C_{(pp)_{{^3}{\rm P}_1}}(p_0) = C_{(pp)_{{^3}{\rm P}_0}}(p_0)
 &=&  \frac{g^2_{\pi NN}}{4p^2_0}\,{\ell n}\Big(1 + 
  \frac{4p^2_0}{m^2_{\pi}}\Big) + (3 - 4\alpha_D)^2\,
\frac{g^2_{\pi NN}}{12 p^2_0}\,{\ell n}\Big(1 +
  \frac{4p^2_0}{m^2_{\eta}}\Big) = \nonumber\\
&=& 3.05\times 10^{-4}\,{\rm MeV}^{-2},
\end{eqnarray}
The cross section for the reaction $pp \to p\Lambda^0K^+$ is equal to
\begin{eqnarray}\label{label2.10}
  \sigma^{pp\to p\Lambda^0K^+}(\varepsilon) &=& \sigma^{(pp)_{{^3}{\rm
        P}_0}\to (p\Lambda^0)_{{^1}{\rm S}_0}K^+}(\varepsilon) +
  \sigma^{(pp)_{{^3}{\rm P}_1}\to (p\Lambda^0)_{{^3}{\rm
        S}_1}K^+}(\varepsilon) =\nonumber\\
  &=& (1.50\,\varepsilon^2 + 3.70\,\varepsilon^2)\,
\Omega_{p\Lambda^0 K^+}(\varepsilon)\,{\rm n b} = 3.49\,\varepsilon^2\,{\rm n b},
\end{eqnarray}
where $\Omega_{p\Lambda^0 K^+}(\varepsilon) = 0.67$ is calculated at
$\varepsilon = 6.68\,{\rm MeV}$ (see Appendix A).  The theoretical
value (\ref{label2.10}) agrees with the experimental data
$\sigma^{pp\to p\Lambda^0K^+}(\varepsilon)_{\exp} =
3.68\,\varepsilon^2\,{\rm n b}$ \cite{JB98} within an accuracy better
than $6\,\%$.

Now we take into account the contribution of the $\rho^0$\,-- and
$\omega^0$\,--\,mesons. The effective Lagrangian of the $pp \to
p\Lambda^0K^+$ transition, caused by the vector\,--\,meson exchanges,
is equal to\,\footnote{Since we compare are results with those by
  F\"aldt and Wilkin \cite{CW97}, for the calculation of the
  contribution of vector--meson exchanges we follow F\"aldt and Wilkin
  \cite{CW97} and treat vector mesons as Yang--Mills fields
  \cite{HP73,GG69}. The interactions of vector mesons with hadronic
  fields are defined by a minimal extension \cite{HP73,GG69}. For the
  derivation of the Lagrangian of the interactions of baryons with
  vector mesons one can use the Lagrangian ({\rm D}.4) of
  Ref.\cite{IV4} with the replacements $g_{\pi NN} \to g_{\rho}/2$,
  $\alpha_D \to 0$, $i\gamma^5 \to \gamma^{\mu}$, $\pi(x) \to
  \rho_{\mu}(x)$ and $\eta(x) \to \omega^0_{\mu}(x)/\sqrt{3}$.}
\begin{eqnarray}\label{label2.11}
  \hspace{-0.3in}\delta {\cal L}^{pp \to p\Lambda^0K^+}_{\rm eff}(x) &=& 
-\, \frac{1}{2\sqrt{3}}\,\frac{(3 - 2\alpha_D)\,g_{\pi NN}\,
    g^2_{\rho}}{m^2_{\rho} + 2 m_N T_N}\,\frac{1}{m_N + m_{\Lambda^0} + m_K}\nonumber\\
  \hspace{-0.3in}&&\times\,
  i\,[\bar{\Lambda}^0(x)\gamma_{\mu}\gamma^5p(x)][\bar{p}(x) \gamma^{\mu}p(x)]\,K^{+\dagger}(x).
\end{eqnarray}
By a Fierz transformation we reduce the effective interaction to the
form
\begin{eqnarray}\label{label2.12}
  &&i\,[\bar{\Lambda}^0(x)\gamma_{\mu}\gamma^5p(x)][\bar{p}(x) \gamma^{\mu}p(x)]
 \to  -\,[\bar{\Lambda}^0(x)i\gamma^5 p^c(x)][\bar{p^c}(x)p(x)]\nonumber\\
  &&\hspace{2.15in}-\,\frac{1}{2}\,
[\bar{\Lambda}^0(x)i\gamma_{\mu}p^c(x)][\bar{p^c}(x)
  \gamma^{\mu}\gamma^5 p(x)] + \ldots\,.
\end{eqnarray}
The contributions of the $\rho^0$\,-- and $\omega^0$\,--\,meson
exchanges to the coefficients $C^{(pp)_{{^3}{\rm
      P}_0}}_{(p\Lambda^0)_{{^1}{\rm S}_0}K^+}$ and $C^{(pp)_{{^3}{\rm
      P}_1}}_{(p\Lambda^0)_{{^3}{\rm S}_1}K^+}$ are equal to
\begin{eqnarray}\label{label2.13}
  \hspace{-0.5in}&&\delta C^{(pp)_{{^3}{\rm
        P}_0}}_{(p\Lambda^0)_{{^1}{\rm S}_0}K^+} = -\frac{2}{\sqrt{3}}
  \frac{(3 - 2\alpha_D)g_{\pi NN} 
    g^2_{\rho}}{m^2_{\rho} + 2 m_N T_N}\frac{1}{m_N + m_{\Lambda^0} + m_K}
  = - 0.30 \times 10^{-6}\,{\rm MeV}^{-3},\nonumber\\
  \hspace{-0.5in}&&\delta C^{(pp)_{{^3}{\rm
        P}_1}}_{(p\Lambda^0)_{{^3}{\rm S}_1}K^+} = +\frac{1}{\sqrt{3}}
  \frac{(3 - 2\alpha_D)g_{\pi NN}
    g^2_{\rho}}{m^2_{\rho} + 2 m_N T_N}\frac{1}{m_N + m_{\Lambda^0} + m_K}
  = + 0.15\times 10^{-6}\,{\rm MeV}^{-3}.
\end{eqnarray}
For the numerical calculation we set $m_{\rho} = m_{\omega} =
780\,{\rm MeV}$ \cite{PDG04}. The contributions of the $\rho^0$\,--
and $\omega^0$\,--\,meson exchanges to the effective coupling constants
$C_{(pp)_{{^3}{\rm P}_0}}(p_0)$ and $C_{(pp)_{{^3}{\rm P}_1}}(p_0)$
amount to
\begin{eqnarray}\label{label2.14}
  \delta C_{(pp)_{{^3}{\rm P}_0}} = -\,\frac{g^2_{\rho}}{2p^2_0}\,
  {\ell n}\Big(1 + \frac{4 p^2_0}{m^2_{\rho}}\Big) =
  -\, 0.43\times 10^{-4}\,{\rm MeV}^{-2},\nonumber\\
  \delta C_{(pp)_{{^3}{\rm P}_1}} = +\,\frac{g^2_{\rho}}{4p^2_0}\,
  {\ell n}\Big(1 + \frac{4 p^2_0}{m^2_{\rho}}\Big) =
  +\,0.22\times 10^{-4}\,{\rm MeV}^{-2}.
\end{eqnarray}
This gives $|f_{(pp)_{{^3}{\rm P}_0}}(p_0)| = 0.16$ and
$|f_{(pp)_{{^3}{\rm P}_1}}(p_0)| = 0.22$. The cross section for the
reaction $pp \to p\Lambda^0K^+$ with the $\rho^0$\,-- and
$\omega^0$\,--\,meson exchanges is
\begin{eqnarray}\label{label2.15}
  \sigma^{pp\to p\Lambda^0K^+}(\varepsilon) &=& \sigma^{(pp)_{{^3}{\rm
        P}_0}\to (p\Lambda^0)_{{^1}{\rm S}_0}K^+}(\varepsilon) +
  \sigma^{(pp)_{{^3}{\rm P}_1}\to (p\Lambda^0)_{{^3}{\rm
        S}_1}K^+}(\varepsilon) =\nonumber\\
  &=& (2.19\,\varepsilon^2 + 2.55\,\varepsilon^2)\,\Omega_{p\Lambda^0 K^+}(\varepsilon)\,{\rm n b} = 3.18\,
  \varepsilon^2\,{\rm n b}.
\end{eqnarray}
Thus, the $\rho^0$\,-- and $\omega^0$\,--\,meson exchanges decrease
the value of the cross section for the reaction $pp\to p\Lambda^0K^+$
by about $9\,\%$ and lead to the agreement with the experimental data
$\sigma^{pp\to p\Lambda^0K^+}(\varepsilon)_{\exp} =
3.68\,\varepsilon^2\,{\rm n b}$ \cite{JB98} with an accuracy about
$16\,\%$.

\section{Cross section for the reaction $pn\to n \Lambda^0K^+$  near
  threshold}
\setcounter{equation}{0}

The $pn$ pair in the reaction $pn \to n\Lambda^0K^+$ can interact both
in the isospin--triplet $(pn)_{I = 1}$ state and isospin--singlet
$(pn)_{I = 0}$ state. According to the {\it generalised Pauli
  exclusion principle}, the $pn$ pair with isospin $I = 1$ can be in
the ${^3}{\rm P}_0$ and ${^3}{\rm P}_1$ state with the $n\Lambda^0$
pair in the ${^1}{\rm S}_0$ and ${^3}{\rm S}_1$ state, respectively,
whereas the $pn$ pair in the state with isospin $I =0$ can be in the
${^1}{\rm P}_1$ state only with the $n\Lambda^0$ pair in the ${^3}{\rm
  S}_1$ state.

A complete set of Feynman diagrams for the $pn \to n\Lambda^0 K^+$
transition near threshold of the final state is depicted in Fig.3 and
Fig.4.
\begin{figure}
\centering 
\psfrag{K+}{$K^+$} 
\psfrag{K-}{$K^-$}
\psfrag{K0}{$K^0$}
\psfrag{L0}{$\Lambda^0$} 
\psfrag{p0}{$\pi^0$}
\psfrag{p+}{$\pi^+$}
\psfrag{S0}{$\Sigma^0$}
\psfrag{S-}{$\Sigma^-$}
\psfrag{p}{$p$} 
\psfrag{n}{$n$} 
\psfrag{e}{$\eta$}
\psfrag{s}{$\sigma$}
\psfrag{d}{$\delta^0$}
\psfrag{k+}{$\kappa^+$}
\psfrag{k-}{$\kappa^-$}
\includegraphics[height= 0.70\textheight]{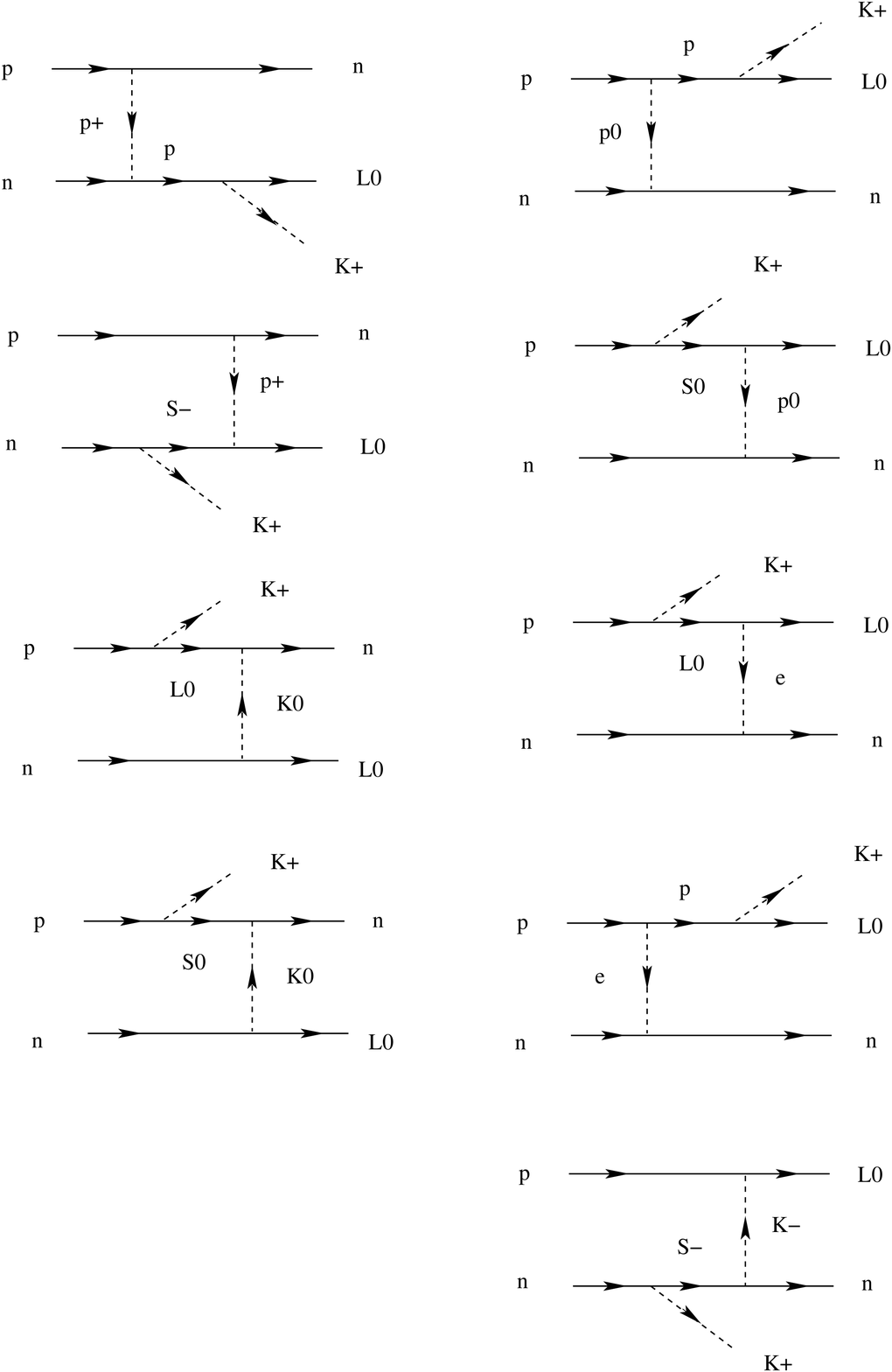}
\caption{The set of Feynman diagrams contributing to the effective
  coupling constants of the $p n \to n
  \Lambda^0K^+$ transition in the one--pseudoscalar meson exchange
  approximation.}
\end{figure}
\begin{figure}
\centering 
\psfrag{K+}{$K^+$}  
\psfrag{K0}{$K^0$}
\psfrag{L0}{$\Lambda^0$} 
\psfrag{p0}{$\pi^0$}
\psfrag{p-}{$\pi^-$}
\psfrag{e}{$\eta$}
\psfrag{p}{$p$}
\psfrag{n}{$n$}
\psfrag{d-}{$\delta^-$}
\psfrag{k-}{$\kappa^-$}
\psfrag{k0}{$\kappa^0$}
\includegraphics[height= 0.30\textheight]{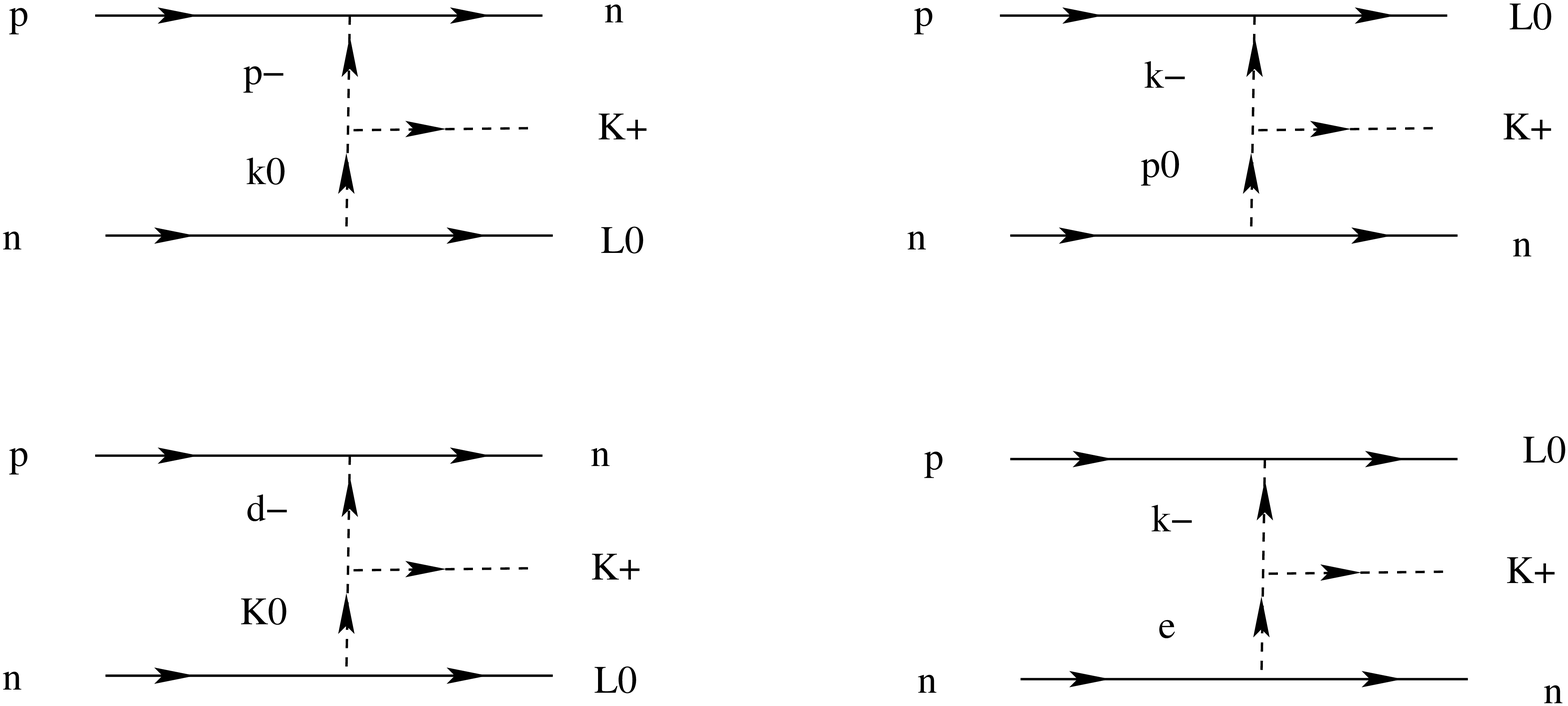}
\caption{The set of Feynman diagrams contributing to the effective
  coupling constants of the $ p n \to n
  \Lambda^0 K^+$ transition in the one--scalar meson exchange
  approximation.}
\end{figure}
The wave functions of the initial and final states we take in the form
\begin{eqnarray}\label{label3.1}
  |p(\vec{p}_1,\sigma_1)n(\vec{p}_2,\sigma_2)\rangle &=&
 a^{\dagger}_p(\vec{p}_1,\sigma_1)a^{\dagger}_n(\vec{p}_2,\sigma_2)|0\rangle,\nonumber\\
  |n(\vec{k}_n,\sigma_n)\Lambda^0(\vec{k}_{\Lambda},
\sigma_{\Lambda})K^+(\vec{Q}_K)\rangle &=& a^{\dagger}_n(\vec{k}_n,\sigma_n)
a^{\dagger}_{\Lambda}
(\vec{k}_{\Lambda},\sigma_{\Lambda})c^{\dagger}_{K^+}(\vec{Q}_K)|0\rangle,
\end{eqnarray}
The matrix element of the $pn \to n\Lambda^0K^+$ transition, defined
by the Feynman diagrams in Fig.3 and Fig.4, is described by the
effective local Lagrangian of the transition $pn \to n\Lambda^0K^+$:
\begin{eqnarray}\label{label3.2}
  {\cal L}^{pn \to n\Lambda^0K^+}_{\rm eff}(x) &=& (A^{pn}_{n\Lambda^0K^+}
  [\bar{\Lambda}^0(x)n(x)][\bar{n}(x)i\gamma^5 p(x)] + 
  B^{pn}_{n\Lambda^0K^+}[\bar{\Lambda}^0(x)p(x)][\bar{n}(x)i\gamma^5 n(x)]
  \nonumber\\
  &+& C^{pn}_{n\Lambda^0K^+}[\bar{\Lambda}^0(x)i\gamma^5 n(x)][\bar{n}(x) p(x)]
  + D^{pn}_{n\Lambda^0K^+}[\bar{\Lambda}^0(x)i\gamma^5 p(x)][\bar{n}(x) n(x)]),
  \nonumber\\
  &&\times\,K^{+\dagger}(x),
\end{eqnarray}
where the effective coupling constants $A^{pn}_{n\Lambda^0K^+}$,
$B^{pn}_{n\Lambda^0K^+}$, $C^{pn}_{n\Lambda^0K^+}$ and
$D^{pn}_{n\Lambda^0K^+}$ are equal to
\begin{eqnarray}\label{label3.3}
  A^{pn}_{n\Lambda^0K^+}&=& -\,\frac{2}{\sqrt{3}}\,\frac{(3 -2 \alpha_D)
    \,g^3_{\pi NN}}{m^2_{\pi} + 2 m_N T_N}\,\frac{1}{m_N + 
m_{\Lambda^0} + m_K}\nonumber\\
  &&+\,\frac{4}{\sqrt{3}}\,\frac{\alpha_D\,(2\alpha_D - 1)\,
g^3_{\pi NN}}{m^2_{\pi} + 2 m_N T_N}\,\frac{m_{\Sigma} + 
m_K - m_N}{m^2_{\Sigma} + 2m_K T_N - (m_N - m_K)^2}\nonumber\\
  &&+\,\frac{1}{\sqrt{3} g^2_A}\,\frac{1}{m_N}\,
  \frac{(3 - 2\alpha_D)\,g^3_{\pi NN}}{m^2_{\pi} + 2 m_N T_N} =
 0.23\times 10^{-6}\,{\rm MeV}^{-3},\nonumber\\
  B^{pn}_{n\Lambda^0K^+}&=&+\,\frac{1}{\sqrt{3}}\,\frac{(3 -2 \alpha_D)
    \,g^3_{\pi NN}}{m^2_{\pi} + 2 m_N T_N}\,
\frac{1}{m_N + m_{\Lambda^0} + m_K}\nonumber\\
  &&-\,\frac{2}{\sqrt{3}}\,\frac{\alpha_D\,(2\alpha_D - 1)
\,g^3_{\pi NN}}{m^2_{\pi} + 2 m_N T_N}\,
\frac{m_{\Sigma} + m_K - m_N}{m^2_{\Sigma} + 2m_K T_N - (m_N - m_K)^2}
  \nonumber\\
  &&+\,\frac{2}{3\sqrt{3}}\,\frac{\alpha_D\,(3 - 2\alpha_D)
    (3 - 4 \alpha_D)\,g^3_{\pi NN}}{m^2_{\eta} + 2 m_N T_N}\,
\frac{m_{\Lambda^0} + m_K - m_N}{m^2_{\Lambda} + 
2 m_K T_N - (m_N - m_K)^2}\nonumber\\
  &&-\,\frac{1}{3\sqrt{3}}\,\frac{(3 - 2\alpha_D)(3 - 
4\alpha_D)^2\,g^3_{\pi NN}}{m^2_{\eta} + 2 m_N T_N}\,
\frac{1}{m_N + m_{\Lambda^0} + m_K}\nonumber\\
  &&+\,\frac{1}{2\sqrt{3} g^2_A}\,\frac{1}{m_N}\,
  \frac{(3 - 2\alpha_D)\,g^3_{\pi NN}}{m^2_{\pi} + 2 m_N T_N} + 
  \frac{1}{6\sqrt{3} g^2_A}\,\frac{1}{m_N}\,\frac{(3 - 2\alpha_D)(3 -
 4 \alpha_D)\,g^3_{\pi NN}}{m^2_{\eta} + 2 m_N T_N}=\nonumber\\
  &&= 2.52\times 10^{-6}\,{\rm MeV}^{-3},\nonumber\\
  C^{pn}_{n\Lambda^0K^+}&=&-\,\frac{1}{3\sqrt{3}}\,
  \frac{(3 - 2\alpha_D)^3\,g^3_{\pi NN}}{m^2_K +
 2 m_{\Lambda^0}T_N - (m_{\Lambda^0} - m_N)^2}\,
\frac{m_{\Lambda^0} + m_K - m_N}{m^2_{\Lambda} + 2 m_K T_N - (m_N - m_K)^2}
  \nonumber\\
  &&+\,\frac{1}{\sqrt{3}}\,
  \frac{(3 - 2\alpha_D)(2\alpha_D - 1)^2\,g^3_{\pi NN}}{m^2_K + 
2 m_{\Lambda^0}T_N - (m_{\Lambda^0} - m_N)^2}\,\frac{m_{\Sigma} + 
m_K - m_N}{m^2_{\Sigma} + 
2 m_K T_N - (m_N - m_K)^2}\nonumber\\
  &&+\,\frac{1}{\sqrt{3} g^2_A}\,\frac{1}{m_N}\,
  \frac{(3 - 2\alpha_D)\,g^3_{\pi NN}}{m^2_K + 2 m_{\Lambda^0} T_N - 
(m_{\Lambda^0} - m_N)^2} = 
  0.51\times 10^{-6}\,{\rm MeV}^{-3},\nonumber\\
 D^{pn}_{n\Lambda^0K^+} &=& \frac{2}{\sqrt{3}}\,
  \frac{(3 - 2\alpha_D)(2\alpha_D - 1)^2\,g^3_{\pi NN}}{m^2_K + 
2 m_{\Lambda^0}T_N - (m_{\Lambda^0} - m_N)^2}\,\frac{m_{\Sigma} + 
m_K - m_N}{m^2_{\Sigma} + 
2 m_K T_N - (m_N - m_K)^2} =\nonumber\\
&=& 0.17\times 10^{-6}\,{\rm MeV}^{-3}.
\end{eqnarray}
Since at low energies the $n\Lambda^0$ pair can be produced in the
${^1}{\rm S}_0$ and ${^3}{\rm S}_1$ state only, we have to extract
these interactions from the effective Lagrangian Eq.(\ref{label3.2})
by a Fierz transformation. We get
\begin{eqnarray}\label{label3.4}
  \hspace{-0.3in} &&{[\bar{\Lambda}^0(x)n(x)][\bar{n}(x)i\gamma^5 p(x)]} = -\,\frac{1}{4}\,
  [\bar{\Lambda}^0(x)i\gamma^5 n^c(x)][\bar{n^c}(x)p(x)]\nonumber\\
  \hspace{-0.3in} &&+\,\frac{1}{4}\,[\bar{\Lambda}^0(x)i\gamma_{\mu}n^c(x)]
  [\bar{n^c}(x)\gamma^{\mu}\gamma^5 p(x)] -\,\frac{1}{8}\,
  [\bar{\Lambda}^0(x)i\sigma_{\mu\nu} n^c(x)][\bar{n^c}(x)\sigma^{\mu\nu}\gamma^5 p(x)] + \ldots,
  \nonumber\\
  \hspace{-0.3in} &&{[\bar{\Lambda}^0(x)p(x)][\bar{n}(x)i\gamma^5 n(x)]} = -\,\frac{1}{4}\,
  [\bar{\Lambda}^0(x)i\gamma^5 n^c(x)][\bar{n^c}(x)p(x)]\nonumber\\
  \hspace{-0.3in} &&+\,\frac{1}{4}\,[\bar{\Lambda}^0(x)i\gamma_{\mu}n^c(x)]
  [\bar{n^c}(x)\gamma^{\mu}\gamma^5 p(x)] + \,\frac{1}{8}\,
  [\bar{\Lambda}^0(x)i\sigma_{\mu\nu} n^c(x)][\bar{n^c}(x)\sigma^{\mu\nu}\gamma^5 p(x)] + \ldots,
  \nonumber\\
  \hspace{-0.3in} &&{[\bar{\Lambda}^0(x)i\gamma^5 n(x)][\bar{n}(x)p(x)]} = -\,\frac{1}{4}\,
  [\bar{\Lambda}^0(x)i\gamma^5 n^c(x)][\bar{n^c}(x)p(x)]\nonumber\\
  \hspace{-0.3in} &&-\,\frac{1}{4}\,[\bar{\Lambda}^0(x)i\gamma_{\mu}n^c(x)]
  [\bar{n^c}(x)\gamma^{\mu}\gamma^5 p(x)] -\,\frac{1}{8}\,
  [\bar{\Lambda}^0(x)i\sigma_{\mu\nu} n^c(x)][\bar{n^c}(x)\sigma^{\mu\nu}\gamma^5 p(x)] + \ldots,
  \nonumber\\
  \hspace{-0.3in} &&{[\bar{\Lambda}^0(x)i\gamma^5 p(x)][\bar{n}(x) n(x)]} = -\,\frac{1}{4}\,
  [\bar{\Lambda}^0(x)i\gamma^5 n^c(x)][\bar{n^c}(x)p(x)]\nonumber\\
  \hspace{-0.3in} &&-\,\frac{1}{4}\,[\bar{\Lambda}^0(x)i\gamma_{\mu}n^c(x)]
  [\bar{n^c}(x)\gamma^{\mu}\gamma^5 p(x)] + \,\frac{1}{8}\,
  [\bar{\Lambda}^0(x)i\sigma_{\mu\nu} n^c(x)][\bar{n^c}(x)\sigma^{\mu\nu}\gamma^5 p(x)] + \ldots.
\end{eqnarray}
Hence, the effective Lagrangian of the $pn\to n\Lambda^0K^+$
transition is equal to
\begin{eqnarray}\label{label3.5}
{\cal L}^{pn \to n\Lambda^0K^+}_{\rm eff}(x) &=& -\,\frac{1}{2}\,
C^{(pn)_{{^3}{\rm P}_0}}_{(n\Lambda^0)_{{^1}{\rm S}_0}K^+}\,
[\bar{\Lambda}^0(x)i\gamma^5 n^c(x)][\bar{n^c}(x)p(x)]\,K^{+\dagger}(x)\nonumber\\
&& + \,\frac{1}{2}\,
C^{(pn)_{{^3}{\rm P}_1}}_{(n\Lambda^0)_{{^3}{\rm S}_1}K^+}\,
[\bar{\Lambda}^0(x)i\gamma_{\mu} n^c(x)][\bar{n^c}(x)\gamma^{\mu}\gamma^5 p(x)]
\,K^{+\dagger}(x)\nonumber\\
&& - \,\frac{1}{4}\,
C^{(pn)_{{^1}{\rm P}_1}}_{(n\Lambda^0)_{{^3}{\rm S}_1}K^+}\,
[\bar{\Lambda}^0(x)i\sigma_{\mu\nu} n^c(x)][\bar{n^c}(x)\sigma^{\mu\nu}\gamma^5 p(x)]
\,K^{+\dagger}(x),
\end{eqnarray}
where the first and the second terms describe the production of the
$n\Lambda^0$ pair in the ${^1}{\rm S}_0$ and ${^3}{\rm S}_1$ state by
the $pn$ pair in the isospin--triplet and the ${^3}{\rm P}_0$ and
${^3}{\rm P}_1$ states, respectively, the third term corresponds to
the interaction of the $pn$ pair in the isospin--singlet and the
${^1}{\rm P}_1$ state. At threshold for the $pn$ pair in the ${^1}{\rm
  P}_1$ state with isospin $I = 0$ the $n\Lambda^0$ pair produces
itself in the ${^3}{\rm S}_1$ state.

The effective coupling constants are equal to
\begin{eqnarray}\label{label3.6}
  C^{(pn)_{{^3}{\rm P}_0}}_{(n\Lambda^0)_{{^1}{\rm S}_0}K^+} &=&  
  \frac{1}{2}\,(A^{pn}_{n\Lambda^0K^+} +  B^{pn}_{n\Lambda^0K^+} + 
  C^{pn}_{n\Lambda^0K^+} +  D^{pn}_{n\Lambda^0K^+}) = \nonumber\\
  &=&+\,1.72\times 10^{-6}\,
  {\rm MeV}^{-3},\nonumber\\
  C^{(pn)_{{^3}{\rm P}_1}}_{(n\Lambda^0)_{{^3}{\rm S}_1}K^+} &=& 
  \frac{1}{2}\,(A^{pn}_{n\Lambda^0K^+} +  B^{pn}_{n\Lambda^0K^+} - 
  C^{pn}_{n\Lambda^0K^+} -  D^{pn}_{n\Lambda^0K^+}) = \nonumber\\
  &=& +\,1.04\times 10^{-6}\,{\rm MeV}^{-3},\nonumber\\
C^{(pn)_{{^1}{\rm P}_1}}_{(n\Lambda^0)_{{^3}{\rm S}_1}K^+} &=&\frac{1}{2}\,
(A^{pn}_{n\Lambda^0K^+} -  B^{pn}_{n\Lambda^0K^+} + 
  C^{pn}_{n\Lambda^0K^+} -  D^{pn}_{n\Lambda^0K^+}) = \nonumber\\
  &=&-\,1.95\times 10^{-6}\,
  {\rm MeV}^{-3},
\end{eqnarray}
The total cross section for the reaction $pn \to
n\Lambda^0K^+$ is given by
\begin{eqnarray}\label{label3.7}
  \sigma^{pn\to n\Lambda^0K^+}(\varepsilon) =
\frac{1}{2}\,( \sigma^{(pn)_{I = 1} \to n\Lambda^0K^+}(\varepsilon) + \sigma^{(pn)_{I = 0} 
\to n\Lambda^0K^+}(\varepsilon)),
\end{eqnarray}
where the cross sections $\sigma^{(pn)_{I = 1} \to
  n\Lambda^0K^+}(\varepsilon)$ and $ \sigma^{(pn)_{I = 0} \to
  n\Lambda^0K^+}(\varepsilon)$ are defined by
\begin{eqnarray}\label{label3.8}
 \hspace{-0.3in} &&\sigma^{(pn)_{I = 1}\to n\Lambda^0K^+}(\varepsilon) = 2\,(\sigma^{(pn)_{{^3}{\rm
        P}_0}\to (n\Lambda^0)_{{^1}{\rm S}_0}K^+}(\varepsilon) +
  \sigma^{(pn)_{{^3}{\rm P}_1}\to (n\Lambda^0)_{{^3}{\rm
        S}_1}K^+}(\varepsilon)) =\nonumber\\
 \hspace{-0.3in}  &&\hspace{1in} = \frac{1}{64\pi^2}\frac{\varepsilon^2}{m_K}\frac{p_0}{E_0}\,
  \Big(\frac{m_Km_{\Lambda^0}m_N}{m_K + m_{\Lambda^0} +
    m_N}\Big)^{3/2}\nonumber\\
 \hspace{-0.3in}  &&
  \times \,\Big(\frac{1}{3}\,
  \Big|C^{(pn)_{{^3}{\rm P}_0}}_{(n\Lambda^0)_{{^1}{\rm S}_0}K^+}\Big|^2
  \Big|f_{(pn)_{{^3}{\rm
        P}_0}}(p_0)\Big|^2 + \frac{2}{3}\,
\Big|C^{(pn)_{{^3}{\rm P}_1}}_{(p\Lambda^0)_{{^3}{\rm S}_1}K^+}\Big|^2
\Big|f_{(pn)_{{^3}{\rm
        P}_1}}(p_0)\Big|^2\Big)\,\Omega_{n\Lambda^0 K^+}(\varepsilon),\nonumber\\
\hspace{-0.3in} &&\sigma^{(pn)_{I = 0}\to n\Lambda^0K^+}(\varepsilon) = 2\,\sigma^{(pn)_{{^1}{\rm
        P}_1}\to (n\Lambda^0)_{{^1}{\rm S}_0}K^+}(\varepsilon) =  \nonumber\\
  \hspace{-0.3in} &&= 
\frac{1}{64\pi^2}\frac{\varepsilon^2}{m_K}\frac{p_0}{E_0}\,
  \Big(\frac{m_Km_{\Lambda^0}m_N}{m_K + m_{\Lambda^0} +
    m_N}\Big)^{3/2}\,\frac{1}{3}\,
  \Big|C^{(pn)_{{^1}{\rm P}_1}}_{(n\Lambda^0)_{{^3}{\rm S}_1}K^+}\Big|^2
  \Big|f_{(pn)_{{^1}{\rm
        P}_1}}(p_0)\Big|^2\,\Omega_{n\Lambda^0 K^+}(\varepsilon).
\end{eqnarray}
The function $\Omega_{n\Lambda^0 K^+}(\varepsilon)$ takes into account
the final--state interaction in the $n\Lambda^0$ channel (see Appendix
A). The amplitude $f_{(pn)_{{^1}{\rm P}_1}}(p_0)$ describes the
rescattering of the $pn$ pair in the isospin--singlet and ${^1}{\rm
  P}_1$ state.

Since the amplitudes of the $pn$ rescattering in the ${^3}{\rm P}_0$
and ${^3}{\rm P}_1$ state, $f_{(pn)_{{^3}{\rm P}_0}}(p_0)$ and
$f_{(pn)_{{^3}{\rm P}_1}}(p_0)$, are equal to the amplitudes of the
$pp$ rescattering, the value of the cross section for the reaction
$(pn)_{I = 1} \to n\Lambda^0K^+$ is equal to
\begin{eqnarray}\label{label3.9}
  \sigma^{(pn)_{I = 1}\to n\Lambda^0K^+}(\varepsilon) = 
(1.71\,\varepsilon^2 + 3.21\,\varepsilon^2)\,\Omega_{n\Lambda^0 K^+}(\varepsilon)\,{\rm n b} = 
3.30\,\varepsilon^2\,{\rm n b},
\end{eqnarray}
where $\Omega_{n\Lambda^0 K^+}(\varepsilon) = 0.67$ for $\varepsilon =
6.68\,{\rm MeV}$ (see Appendix A).

For the cross sections of the reactions $(pn)_{I = 1}\to
n\Lambda^0K^+$ and $pp \to p \Lambda^0K^+$, defined by
Eqs.(\ref{label2.10}) and (\ref{label3.9}), we get the ratio
\begin{eqnarray}\label{label3.10}
  R^{(I = 1)} = \frac{\sigma^{(pn)_{I = 1}\to n\Lambda^0K^+}(\varepsilon)}{
\sigma^{pp\to p\Lambda^0K^+}(\varepsilon)} \simeq 1,
\end{eqnarray}
agreeing well with isospin--invariance of strong $pN$ interactions.

According to \cite{IV4}, the result of the calculation of the
amplitude $f_{(pn)_{{^1}{\rm P}_1}}(p_0)$ is
\begin{eqnarray}\label{label3.11}
  |f_{(pn)_{{^1}{\rm P}_1}}(p_0)|  =  \Big|\Big\{1 - 
  \frac{ C_{(pn)_{{^1}{\rm P}_1}}(p_0)}{24\pi^2}\,
  \frac{p^3_0}{E_N}\,\Big[{\ell n}\Big(\frac{E_N + p_0}{E_N - p_0}\Big)
  -\,i\,\pi\Big]\Big\}^{-1}\Big| = 0.49,
\end{eqnarray}
where the coupling constant $C_{(pn)_{{^1}{\rm P}_1}}(p_0)$ is defined
by the $\pi$\,-- and $\eta$\,--\,meson exchanges and is equal to
$C_{(pn)_{{^1}{\rm P}_1}}(p_0) = C_{(pn)_{{^3}{\rm P}_1}}(p_0) =
3.05\times 10^{-4}\,{\rm MeV}^{-2}$ \cite{IV4}. The cross section for
the reaction $(pn)_{I = 0} \to n\Lambda^0K^+$ amounts to
\begin{eqnarray}\label{label3.12}
  \sigma^{(pn)_{I = 0}\to n\Lambda^0K^+}(\varepsilon) = 
  23.50\,\varepsilon^2\,\Omega_{n\Lambda^0K^+}(\varepsilon)\,{\rm n b} = 15.75\,\varepsilon^2\,{\rm n b}.
\end{eqnarray}
For the total cross section for the reaction $pn \to n\Lambda^0K^+$ we get
\begin{eqnarray}\label{label3.13}
  \sigma^{pn\to n\Lambda^0K^+}(\varepsilon) &=&
  \frac{1}{2}(\sigma^{(pn)_{I = 1} \to n\Lambda^0K^+}(\varepsilon) + \sigma^{(pn)_{I = 0} 
    \to n\Lambda^0K^+}(\varepsilon)) =  \nonumber\\
  &=& \Delta \sigma^{pn \to n\Lambda^0K^+}(\varepsilon) + \sigma^{(pn)_{I = 1} \to n\Lambda^0K^+}(\varepsilon) = 
  9.52\,\varepsilon^2\,{\rm n b},
\end{eqnarray}
where we have denoted
\begin{eqnarray}\label{label3.14}
  \Delta \sigma^{pn \to n\Lambda^0K^+}(\varepsilon) = 
  \frac{1}{2}(\sigma^{(pn)_{I = 0} 
    \to n\Lambda^0K^+}(\varepsilon) - \sigma^{(pn)_{I = 1} \to n\Lambda^0K^+}(\varepsilon)) =  
  6.22\,\varepsilon^2\,{\rm n b}.
\end{eqnarray}
Now we can take into account the contribution of the vector meson
exchanges. Unlike the reaction $pp \to p\Lambda^0K^+$ the contribution
of the $\rho^0$\,-- and $\omega^0$\,--\,meson exchanges is cancelled
and there are only the contributions of the charged $\rho$\,--\,meson
exchange. The effective Lagrangian, caused by the charged
$\rho$\,--\,meson exchanges, is equal to
\begin{eqnarray}\label{label3.15}
  \hspace{-0.3in}\delta {\cal L}^{pn \to n\Lambda^0K^+}_{\rm eff}(x) &=& 
-\,\frac{1}{2\sqrt{3}}\,\frac{(3 - 2\alpha_D)\,
    g_{\pi N5}\,g^2_{\rho}}{m^2_{\rho} + 2 m_N T_N}\,\frac{1}{m_N +
    m_{\Lambda^0} + m_K}
  \nonumber\\
  \hspace{-0.3in} &&\times\,
  [\bar{\Lambda}^0(x)\,i\,\gamma_{\mu}\gamma^5n(x)][\bar{n}(x) 
  \gamma^{\mu}p(x)]\,K^{+\dagger}(x).
\end{eqnarray}
By a Fierz transformation
\begin{eqnarray}\label{label3.16}
{[\bar{\Lambda}^0(x)i\gamma_{\mu} \gamma^5 n(x)][\bar{n}(x)\gamma^{\mu} p(x)]}
 &=& -\,\frac{1}{4}\,
[\bar{\Lambda}^0(x)i\gamma^5 n^c(x)][\bar{n^c}(x)p(x)]\nonumber\\
&&-\,\frac{1}{4}\,[\bar{\Lambda}^0(x)i\gamma_{\mu}n^c(x)]
[\bar{n^c}(x)\gamma^{\mu}\gamma^5 p(x)] + \ldots
\end{eqnarray}
we obtain the contributions of the charged $\rho$\,--\,meson exchanges
to the effective coupling constants $C^{(pn)_{{^3}{\rm
      P}_0}}_{(n\Lambda^0)_{{^1}{\rm S}_0}K^+}$, $C^{(pn)_{{^3}{\rm
      P}_1}}_{(n\Lambda^0)_{{^3}{\rm S}_1}K^+}$ and $C^{(pn)_{{^1}{\rm
      P}_1}}_{(n\Lambda^0)_{{^3}{\rm S}_1}K^+}$
\begin{eqnarray}\label{label3.17}
  \delta C^{(pn)_{{^3}{\rm P}_0}}_{(n\Lambda^0)_{{^1}{\rm S}_0}K^+} &=&  
  -\,\frac{1}{4\sqrt{3}}\,\frac{(3 - 2\alpha_D)\,
    g_{\pi NN}\,g^2_{\rho}}{m^2_{\rho} + 2 m_N T_N}\,\frac{1}{m_N +
    m_{\Lambda^0} + m_K} = -\,0.04\times 10^{-6}\,
  {\rm MeV}^{-3} ,\nonumber\\
  \delta C^{(pn)_{{^3}{\rm P}_1}}_{(n\Lambda^0)_{{^3}{\rm S}_1}K^+} &=&+\, 
\frac{1}{4\sqrt{3}}\,\frac{(3 - 2\alpha_D)\,
    g_{\pi NN}\,g^2_{\rho}}{m^2_{\rho} + 2 m_N T_N}\,\frac{1}{m_N +
    m_{\Lambda^0} + m_K} = +\,0.04\times 10^{-6}\,
  {\rm MeV}^{-3},\nonumber\\
\delta C^{(pn)_{{^1}{\rm P}_1}}_{(n\Lambda^0)_{{^3}{\rm S}_1}K^+} &=& 0.
\end{eqnarray}
The contributions of the charged $\rho$\,--\,meson exchanges to the
effective coupling constants $C_{(pn)_{{^3}{\rm P}_0}}(p_0)$ and
$C_{(pn)_{{^3}{\rm P}_1}}(p_0)$ amount to
\begin{eqnarray}\label{label3.18}
  \delta C_{(pn)_{{^3}{\rm P}_0}}(p_0) = -\,\frac{g^2_{\rho}}{2p^2_0}\,
  {\ell n}\Big(1 + \frac{4 p^2_0}{m^2_{\rho}}\Big) =
 -\, 0.43\times 10^{-4}\,{\rm MeV}^{-2},\nonumber\\
\delta C_{(pn)_{{^3}{\rm P}_1}}(p_0) = +\,\frac{g^2_{\rho}}{4p^2_0}\,
  {\ell n}\Big(1 + \frac{4 p^2_0}{m^2_{\rho}}\Big) =
  +\,0.22\times 10^{-4}\,{\rm MeV}^{-2}.
\end{eqnarray}
This gives $|f_{(pn)_{{^3}{\rm P}_0}}(p_0)| = 0.16$ and
$|f_{(pn)_{{^3}{\rm P}_1}}(p_0)| = 0.22$. The cross section for the
reaction $(pn)_{I = 1} \to n\Lambda^0K^+$ with the contribution of the
vector\,--\,meson exchanges is
\begin{eqnarray}\label{label3.19}
  \sigma^{(pn)_{I = 1}\to n\Lambda^0K^+}(\varepsilon) = 
  (1.86\,\varepsilon^2 + 2.91\,\varepsilon^2)\,
\Omega_{n\Lambda^0K^+}(\varepsilon)\,{\rm n b} = 3.20\,\varepsilon^2\,{\rm n b}.
\end{eqnarray}
This gives the following value of the ratio $R^{(I = 1)}$:
\begin{eqnarray}\label{label3.20}
  R^{(I = 1)} = \frac{\sigma^{(pn)_{I = 1}\to n\Lambda^0K^+}(\varepsilon)}{
    \sigma^{pp\to p\Lambda^0K^+}(\varepsilon)} \simeq  1,
\end{eqnarray}
which is in agreement with isospin invariance of strong $pN$ interactions.

The vector\,--\,meson exchanges lead to the decrease of the cross
section for the reaction $(pn)_{I = 1}\to n\Lambda^0K^+$ by about
$3\,\%$.  Thus, the total cross section for the reaction $pn \to
n\Lambda^0K^+$ is equal to
\begin{eqnarray}\label{label3.21}
  \sigma^{pn\to n\Lambda^0K^+}(\varepsilon) &=&
  \frac{1}{2}(\sigma^{(pn)_{I = 1} \to n\Lambda^0K^+}(\varepsilon) + \sigma^{(pn)_{I = 0} 
    \to n\Lambda^0K^+}(\varepsilon)) =  \nonumber\\
  &=& \Delta \sigma^{pn \to n\Lambda^0K^+}(\varepsilon) + \sigma^{(pn)_{I = 1} \to n\Lambda^0K^+}(\varepsilon) = 
\nonumber\\
&=& (6.28\,\varepsilon^2 + 3.20\,\varepsilon^2)\,{\rm n b} = 9.48\,
\varepsilon^2\,{\rm n b}.
\end{eqnarray}
Hence, due to the dominant contribution of the $pn$ interaction in the
isospin--singlet state, the vector\,--\,meson exchanges have
practically no influence on the cross section for the reaction $pn \to
n\Lambda^0K^+$. The decrease of the total cross section for the
reaction $pn \to n \Lambda^0 K^+$, caused by the vector\,--\,meson
exchanges, is about $0.4\,\%$.

\section{Ratio of the cross sections}
\setcounter{equation}{0}

Using the results obtained above we can calculate the ratio of the
cross sections for the reactions $pn \to n\Lambda^0K^+$ and $pp \to
p\Lambda^0K^+$. For the theoretical cross sections, calculated with
the account for vector\,--\,meson exchanges Eq.(\ref{label3.12}) and
Eq.(\ref{label2.15}) and without vector\,--\,meson exchanges
Eq.(\ref{label3.12}) and Eq.(\ref{label2.10}), we get
\begin{eqnarray}\label{label4.1}
  R = \frac{\sigma^{pn\to n\Lambda^0K^+}(\varepsilon)}{\sigma^{pp\to
      p\Lambda^0K^+}(\varepsilon)} \simeq 3.
\end{eqnarray}
The obtained result confirms an enhancement of the $pn$ interaction
with respect to the $pp$ interaction near observed by the
ANKE--Collaboration at COSY \cite{COSY}. In our analysis such an
enhancement is due to the contribution of the $pn$ interaction in the
isospin--singlet state $\Delta \sigma^{pn\to
  n\Lambda^0K^+}(\varepsilon) = \frac{1}{2}(\sigma^{(pn)_{I = 0} \to
  n\Lambda^0K^+}(\varepsilon) - \sigma^{(pn)_{I = 1} \to
  n\Lambda^0K^+}(\varepsilon)) \approx 6.3\,{\rm nb}$.

\section{On the $N(1535)$ and $N(1650)$ resonance exchanges}
\setcounter{equation}{0}

As has been pointed out by Wilkin \cite{CW05}, an important
contribution to the cross sections for the reactions $pN \to
N\Lambda^0K^+$ should come from the exchange with the $N(1535)$
resonance \cite{PDG04a}. This is the $S_{11}(1535)$ resonance with the
quantum numbers $I(J^P) = \frac{1}{2}(\frac{1}{2}^-)$ \cite{PDG04a}.

The inclusion of the $N(1535)$--resonance exchange demands the
knowledge of its interactions with the octets of the ground baryons
and the pseudoscalar mesons. Since the $N(1535)$ resonance is the
member of octet \cite{PDG04b}, the effective Lagrangian of its
interactions with the ground state baryons and the pseudoscalar
mesons reads
\begin{eqnarray}\label{label5.1}
  {\cal L}_{\rm PBN_1}(x) &=& \sqrt{2}\,g_{\pi NN_1}\,\bar{p}(x)n_1(x)\pi^+(x) + 
  \sqrt{2}\,g_{\pi NN_1}\,\bar{n}(x)p_1(x)\pi^-(x)\nonumber\\
  &+& g_{\pi NN_1}\,[\bar{p}(x)p_1(x) - \bar{n}(x)n_1(x)]\,\pi^0(x)\nonumber\\ &+&
  \frac{1}{\sqrt{3}}\,(3 - 4\alpha_{D1})\,g_{\pi NN_1}\,[\bar{p}(x)p_1(x)
  +  \bar{n}(x)n_1(x)]\,\eta(x)\nonumber\\
  &+& \frac{1}{\sqrt{3}}\,(3 - 2\alpha_{D1})\,
  g_{\pi NN_1}\,\bar{\Lambda}^0(x)p_1(x)K^-(x) + \ldots + {\rm h.c.},
\end{eqnarray}
where $N_1(x) = (p_1(x),n_1(x))$ are the local interpolating field of
the resonance $N(1535)$, which we treat as an elementary particle
\cite{MR96}--\cite{IV9} as well as the $\Delta(1232)$ resonance
\cite{AI99,TE05}.

Due to the ambiguity of the parameters of the $N(1535)$ resonance the
numerical value of the contribution of this resonance is rather
ambiguous \cite{PDG04}. Below we use the following parameters: the
mass $m_{N(1535)} = 1520\,{\rm MeV}$, the width $\Gamma_{N(1535)} =
100\,{\rm MeV}$ and the branching ratios $B(N(1535)\to N\pi) = 40\,\%$
and $B(N(1535)\to N\eta) = 35\,\%$.  We leave $25\,\%$ of the total
width for other decay channels such as $N(1535)\to \Delta(1232)\pi$
and so on \cite{PDG04}. For such a choice of the parameters the
coupling constants are equal to $g_{\pi NN_1} = 0.53$ and $\alpha_{D1}
= -\,0.52$.

For the reaction $pp \to p\Lambda^0K^+$ the $N(1535)$ resonance gives
the contribution only to the effective coupling constant
$A^{pp}_{p\Lambda^0K^+}$. We get
\begin{eqnarray}\label{label5.2}
  \hspace{-0.3in}&&(\delta A^{pp}_{p\Lambda^0K^+})_{N(1535)} =   
  \frac{1}{\sqrt{3}}\,\frac{(3 - 2\alpha_{D1})g_{\pi NN}\,
    g^2_{\pi NN_1}}{m_{N_1} - m_{\Lambda^0} - m_K}\nonumber\\
  \hspace{-0.3in}&&\times\,\Big[\frac{1}{m^2_{\pi} + 2m_NT_N}
  + \frac{1}{3}\,\frac{(3 - 4\alpha_{D1})(3 - 4\alpha_{D})}{m^2_{\eta} + 2m_NT_N}\Big] = 
  -\,0.22\times 10^{-6}\,{\rm MeV}^{-3}.
\end{eqnarray}
We can also take into account the contribution of the $N(1650)$
resonance (the $S_{11}(1650)$ resonance) with the quantum numbers
$I(J^P) = \frac{1}{2}(\frac{1}{2}^-)$, which is the octet partner of
the resonances $\Lambda(1800)$ and $\Sigma(1750)$ \cite{PDG04b}. The
coupling constants of the $N(1650)$ resonance with the ground state
baryons and the pseudoscalar mesons are equal to $g_{\pi NN_2} = 0.62$
and $\alpha_{D2} = 1.17$. They are calculated for $m_{N(1650)} =
1640\,{\rm MeV}$, $B(N(1650)\to N\pi) = 70\,\%$ and $B(N(1650)\to
\Lambda^0K) = 10\,\%$ and $\Gamma_{N(1650)} = 170\,{\rm MeV}$
\cite{PDG04}.

The $N(1650)$ resonance as well as the $N(1535)$ resonance gives the
contribution to the effective coupling constant
$A^{pp}_{p\Lambda^0K^+}$. The result is
\begin{eqnarray}\label{label5.3}
  \hspace{-0.3in}&&(\delta A^{pp}_{p\Lambda^0K^+})_{N(1650)} =  \frac{1}{\sqrt{3}}\,
  \frac{(3 - 2\alpha_{D2})g_{\pi NN}\,
    g^2_{\pi NN_2}}{m_{N_2} - m_{\Lambda^0} - m_K}\nonumber\\
  \hspace{-0.3in}&&\times\,\Big[\frac{1}{m^2_{\pi} + 2m_NT_N}
  + \frac{1}{3}\,\frac{(3 - 4\alpha_{D1})(3 - 4\alpha_{D})}{m^2_{\eta} + 2m_NT_N}\Big] = 
  0.08\times 10^{-6}\,{\rm MeV}^{-3}.
\end{eqnarray}
Hence, the total contribution of the resonances $N(1535)$ and
$N(1650)$ is about $8\,\%$.

The values of the parameters of the resonances $N(1535)$ and $N(1650)$
are rather ambiguous. Nevertheless, as we show in Section 7 our choice
of the parameters of the resonances $N(1535)$ and $N(1650)$ fits well
the experimental data on the amplitude and the cross section for the
reaction $\pi^- p \to \Lambda^0 K^0$ near threshold of the
$\Lambda^0K^0$ state.

We would like to accentuate that a cancellation of contributions of
the resonances $N(1535)$ and $N(1650)$ occurs only in the Lagrangian
approach, where the resonances are treated as elementary particles
\cite{MR96}--\cite{IV9}. We cannot claim that such a cancellation
between the contributions of these resonances should be within the
approaches developed by Wilkin {\it et al.}  \cite{CW97,CW05} or by
Sibirtsev {\it et al.}  \cite{KT99,AS05}, where the $N(1535)$ and
$N(1650)$ resonances were described within the relativistic
generalisation of the Breit--Wigner approach \cite{PDG04}.

Thus, treating the the resonances $N(1535)$ and $N(1650)$ as
elementary particles we have shown that the total contribution of the
resonances $N(1535)$ and $N(1650)$ to the matrix elements of the $pN
\to N\Lambda^0K^+$ transitions near threshold of the final state is less
important in comparison with the contribution of the ground state
baryon octet coupled to the octets of the pseudoscalar and scalar
mesons.

\section{Cross section for the reaction $pp \to pp\pi^0$}
\setcounter{equation}{0}

In this Section we show that the same procedure, which we have used
for the analysis of the strangeness production in $pN$ collisions near
threshold, can be applied well to the description of the reaction $pp
\to pp\pi^0$.

Near threshold the reaction $pp \to pp\pi^0$ is defined by a pure
${^3}{\rm P}_0 \to {^1}{\rm S}_0$ transition \cite{HM1}--\cite{NK02}.
The effective Lagrangian responsible for the $(pp)_{{^3}{\rm P}_0} \to
(pp)_{{^1}{\rm S}_0} \pi^0$ transition, calculated in the analogy with
the effective Lagrangians (\ref{label2.2}) and (\ref{label3.5}), is
\begin{eqnarray}\label{label6.1}
  {\cal L}^{(pp\to pp\pi^0)}(x) = \frac{1}{4}\,C^{pp}_{pp\pi^0}(p_0)[\bar{p}(x)p^c(x)]
[\bar{p^c}(x)i\gamma^5p(x)]\pi^0(x).
\end{eqnarray}
The effective coupling constant $C^{pp}_{pp\pi^0}$ is equal to
\begin{eqnarray}\label{label6.2}
  C^{pp}_{pp\pi^0}(p_0) &=& \frac{g^3_{\pi NN}}{m_N}\Big[
  \Big(1 - \frac{1}{g^2_A}\Big)\frac{1}{m^2_{\pi} + p^2_0} + \frac{1}{3}\,
  \frac{(3 - 4\alpha_D)}{m^2_{\eta} + p^2_0}\Big(3 - 4 \alpha_D - \frac{1}{g^2_A}\Big) 
  \Big] = \nonumber\\
  &=& 6.04\times 10^{-6}\,{\rm MeV}^{-3},
\end{eqnarray}
where $p_0 = 362.2\,{\rm MeV}$ is a relative threshold momentum of the
$pp$ pair in the initial state.

For the cross section for the reaction $pp \to pp\pi^0$ we obtain the
following expression
\begin{eqnarray}\label{label6.3}
  \sigma^{pp\to pp\pi^0}(\varepsilon) &=& \frac{1}{192\pi^2}\,\frac{p_0}{E_0}
  \frac{\varepsilon^2}{m_{\pi}}\Big(\frac{m^2_p m_{\pi}}{2 m_N + m_{\pi}}\Big)^{3/2} 
  \vert C^{pp}_{pp\pi^0}(p_0)\vert^2\vert f_{(pp)_{{^3}{\rm P}_0}}(p_0)\vert^2
  \Omega_{pp\pi^0}(\varepsilon) =\nonumber\\
  &=& 0.43\,\varepsilon^2\,\Omega_{pp\pi^0}(\varepsilon)\,{\rm \mu b},
\end{eqnarray}
where $\vert f_{(pp)_{{^3}{\rm P}_0}}(p_0)\vert = 0.38$
Eq.(\ref{label2.8}). The excess energy $\varepsilon$ is related to the
kinetic energy of the proton in the laboratory frame as $\varepsilon =
\frac{1}{2}\,T_N - m_{\pi}$. It is measured in ${\rm MeV}$. The
function $\Omega_{pp\pi^0}(\varepsilon)$ is defined by the phase volume of the
$pp\pi^0$ state. It is given by
\begin{eqnarray}\label{label6.4}
  \Omega_{pp\pi^0}(\varepsilon) = \frac{16}{\pi}\int^1_0d\xi\,\xi^2\,\sqrt{1 - \xi^2}\,
\vert\psi_{pp}(\sqrt{m_N\varepsilon}\,\xi)\vert^2.
\end{eqnarray}
The wave function $\psi_{pp}(k)$, describing the final--state
interaction in the $pp$ channel accounting for the Coulomb repulsion,
is determined by \cite{AI1} (see also \cite{KM58}):
\begin{eqnarray}\label{label6.5}
  \psi_{pp}(k) = e^{\textstyle i\,\delta^{\,\rm e}_{\rm
      pp}(k)}\,\frac{\sin\,\delta^{\,\rm e}_{\rm pp}(k)}{-a^{\rm e}_{\rm
      pp}kC_0(k)} = \frac{\displaystyle C_0(k)}{\displaystyle1
    - \frac{1}{2}\,a^{\rm e}_{\rm pp} r^{\rm e}_{\rm pp}k^2 + \frac{a^{\rm
        e}_{\rm pp}}{r_C}\,h(2 k r_C) + i\,a^{\rm e}_{\rm pp}\,k\,C^2_0(k)},
\end{eqnarray}
where we have denoted \cite{AI1}
\begin{eqnarray}\label{label6.6}
C^2_0(k) &=& \frac{\pi}{k r_C}\,
\frac{1}{\displaystyle e^{\textstyle \pi/k r_C} - 1},\nonumber\\
h(2 k r_C) &=& - \gamma + {\ell n}(2 k r_C) +
\sum^{\infty}_{n=1}\frac{1}{n(1 + 4n^2k^2r^2_C)},\nonumber\\
{\rm ctg}\delta^{\rm e}_{\rm pp}(k) &=& \frac{1}{\displaystyle
C^2_0(k)\,k}\,\Big[ - \frac{1}{a^{\rm e}_{\rm pp}} +  \frac{1}{2}\,r^{\rm
e}_{\rm pp}k^2 -  \frac{1}{r_{\rm C}}\,h(2 k r_{\rm C})\Big],
\end{eqnarray}
where $\gamma = 0.57721\ldots$ is Euler's constant, $r_C = 1/m_N\alpha
= 28.82\,{\rm fm}$ with $\alpha = 1/137.036$ is the fine--structure
constant, $\delta^{\rm e}_{\rm pp}(k)$ is the phase shift of
low--energy elastic pp scattering in terms of the S--wave scattering
length $a^{\rm e}_{\rm pp} = ( - 7.8196\pm 0.0026)\,{\rm fm}$ and the
effective range $r^{\rm e}_{\rm pp} = (2.790\pm 0.014)\,{\rm fm}$
\cite{MN79}. At $\alpha \to 0$ and $k \to 0$ the wave function
$\psi_{pp}(k)$ tends to unity $\psi_{pp}(k) \to 1$ (see \cite{AI1} and
Appendix A).

The theoretical cross section for the reaction $pp\to pp\pi^0$
together with experimental data are depicted in Fig.5. In Table 1 we
adduce numerical values of the theoretical and experimental cross
section for the reaction $pp\to pp\pi^0$.
\begin{figure}[h]
\psfrag{0}{$0$}\psfrag{20}{$20$}\psfrag{40}{$40$}
\psfrag{60}{$60$}\psfrag{80}{$80$}\psfrag{100}{$100$}
\psfrag{a}{$T_N({\rm MeV})\,\rightarrow$}
\psfrag{e}{$\varepsilon({\rm MeV})\,\rightarrow$}
\psfrag{b}{$2.5$}
\psfrag{c}{$5.0$}
\psfrag{d}{$7.5$}
\psfrag{s}{$\sigma(\mu b)$}
\centering
\includegraphics[scale=0.9]{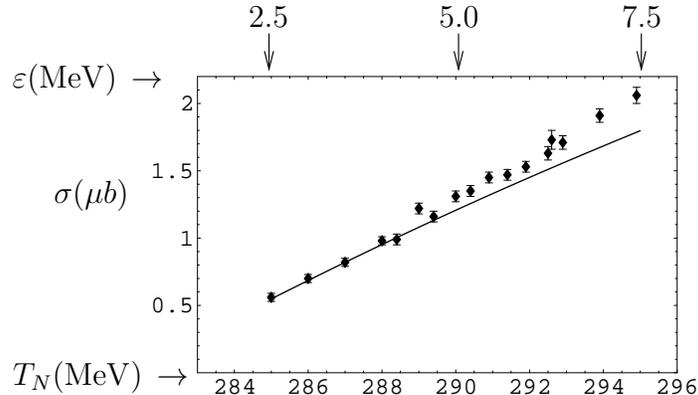}
\caption{The theoretical cross-section for the reaction $pp \to
  pp\pi^0$, defined by the ${^3}{\rm P}_0 \to {^1}{\rm S}_0$
  transition. The experimental data are taken from \cite{HM1}, $T_N$
  is a kinetic energy of the incident proton in the laboratory frame.}
\end{figure}
\begin{table}[h]
\begin{tabular}{|l||r|r|r|r|r|r|r|r|}
  \hline
  $T_N({\rm MeV})$& $285$ & $286$& $287$ & $288$& $289$& $290$& $291$& $292$\\[0.5ex]
  \hline
  $\sigma^{(\exp)}(T_N)$& $0.56(3)$ & $0.70(3) $ & $0.82(3)$& $0.98(3)$ &
  $1.22(4)$ & $1.31(4)$& $1.45(4)$& $1.53(4)$\\
  \hline
  $\sigma^{(\rm th)}(T_N)$& $0.55~~~\,$ & $0.69~~~\,$ & $0.82~~~\,$& $0.95~~~\,$ & $1.08~~~\,$ & 
$1.21~~~\,$& $1.33~~~\,$& $1.45~~~~$\\
  \hline
\end{tabular}
\caption{The theoretical and experimental values of the cross section for the reaction $pp \to 
  pp\pi^0$, caused by the ${^3}{\rm P} \to {^1}{\rm S}_0$ transition. The experimental
  data  are taken from \cite{HM1}. The cross section is measured in ${\mu b}$.}
\end{table}

\noindent The theoretical cross section agrees well with the
experimental data.  This implies that the analysis of the strangeness
production near threshold of $pN$ collisions given above works also
well for the description of the pion--production in the $pp$
collisions.  The cross sections for other channels $pN \to NN\pi$
\cite{NK02} can be calculated in a similar way.

\section{Cross section for the reaction $\pi^0 p \to K^+ \Lambda^0$}
\setcounter{equation}{0}

Within chiral Lagrangian approach the reaction $\pi^0 p \to K^+
\Lambda^0$ has been investigated in \cite{WW97} (see also
\cite{NK99}). Due to isospin invariance the amplitude of the reaction
$\pi^-p \to K^0\Lambda^0$ is related to the amplitude of the reaction
$\pi^0 p \to K^+ \Lambda^0$ as
\begin{eqnarray}\label{label7.1}
M(\pi^-p\to K^0\Lambda^0) = \sqrt{2}\,M(\pi^0p \to K^+ \Lambda^0).
\end{eqnarray}
The relative threshold momentum of the $\pi^0p$ pair is equal to $p_0
= 520\,{\rm MeV}$. At threshold we define the amplitude of the
reaction $\pi^0p \to K^+\Lambda^0$
\begin{eqnarray}\label{label7.2}
  M(\pi^0p \to K^+ \Lambda^0) = 8\pi (m_K + m_{\Lambda^0})\,a_{K^+\Lambda^0}(p_0)
\end{eqnarray}
and the cross section \cite{JJ71}
\begin{eqnarray}\label{label7.3}
  \sigma^{\pi^0p\to K^+\Lambda^0}(p_{\pi^0}) = A_{K^+\Lambda^0}\,p_{K^+\Lambda^0}. 
\end{eqnarray}
Here $p_{K^+\Lambda^0}$ is a relative momentum of the $K^+\Lambda^0$
pair, measured in ${\rm GeV}$, and $A_{K^+\Lambda^0}$ is equal to
\begin{eqnarray}\label{label7.4}
  A_{K^+\Lambda^0} = \frac{4\pi}{p_0}\,\vert a_{K^+\Lambda^0}(p_0)\vert^2.
\end{eqnarray}
For the amplitude $a_{K^+\Lambda^0}$ we obtain the following expression
\begin{eqnarray}\label{label7.5}
  \hspace{-0.3in}&&a_{K^+\Lambda^0}(p_0) = \frac{\sqrt{m_{\Lambda^0}m_N}}{
    4\pi( m_K + m_{\Lambda^0})}\,
  \Big[
  \frac{1}{\sqrt{3}}\,(3 - 2\alpha_D)\,
  \Big(\frac{g^2_{\pi NN}}{m_N + m_K + m_{\Lambda^0}} -
  \frac{g^2_{\pi NN}}{g^2_A}\,\frac{1}{2m_N}\Big)\nonumber\\
  \hspace{-0.3in}&&- \frac{2}{\sqrt{3}}\,\alpha_D(2\alpha_D - 1)\,
  \frac{ g^2_{\pi NN}\,(m_{\Sigma^0} + m_K - m_N)}{m^2_{\Sigma^0} + 
    2m_KT_N - 
    (m_N - m_K)^2} - \frac{1}{\sqrt{3}}\frac{(3 - 2\alpha_{D1})\,
    g^2_{\pi NN_1}}{m_{N_1} - m_K - m_{\Lambda^0}}\nonumber\\
  \hspace{-0.3in}&& -\,\frac{1}{\sqrt{3}}\frac{(3 - 2\alpha_{D2})\,g^2_{\pi NN_2}}{m_{N_2} 
    - m_K - m_{\Lambda^0}} +  \frac{2}{\sqrt{3}}\,\alpha_{D2}(2\alpha_{D2} - 1)\,
  \frac{ g^2_{\pi NN_2}\,(m_{\Sigma^0_2} + m_N - m_K)}{m^2_{\Sigma^0_2} + 
    2m_KT_N - 
    (m_N - m_K)^2}\Big] = \nonumber\\
  \hspace{-0.3in}&&= (0.109 - 0.190_{\Sigma^0} + 0.073_{N_1} - 0.047_{N_2} + 0.005_{\Sigma_2})
  \,{\rm fm} = -\,0.050\,{\rm fm},
\end{eqnarray}
where $m_{\Sigma_2} = 1750\,{\rm MeV}$. We remind that the resonance
$\Sigma(1750)$ is the octet partner of the $N(1650)$--resonance
\cite{PDG04}. One can show that the contribution of the resonance
$\Sigma(1620)$, the octet partner of the resonance $N(1535)$
\cite{PDG04}, is negligible small.

For the constant $A_{K^+\Lambda^0}$ we get the value
\begin{eqnarray}\label{label7.6}
  A_{K^+\Lambda^0} = 0.61\,{\rm m\,b}/{\rm GeV}.
\end{eqnarray}
Due to isospin invariance the constant $A_{K^0\Lambda^0}$ of the
reaction $\pi^-p \to K^0\Lambda^0$ is equal to
\begin{eqnarray}\label{label7.7}
  A_{K^0\Lambda^0} = 2 A_{K^+\Lambda^0} = 1.22\,{\rm m\,b}/{\rm GeV}.
\end{eqnarray}
Near threshold of the reaction $\pi^- p\to K^0\Lambda^0$ the constant
$A_{K^0\Lambda^0}$ defines the cross section \cite{JJ71}
\begin{eqnarray}\label{label7.8}
  \sigma^{\pi^- p \to K^0\Lambda^0}(p_{\ell ab}) = A_{K^0\Lambda^0}\,p_{K^0\Lambda^0}, 
\end{eqnarray}
where $p_{\ell ab}$ is a $\pi^-$--meson momentum in the laboratory
frame and $p_{K^0\Lambda^0}$ is the relative momentum of the
$K^0\Lambda^0$ pair.
\begin{figure}[h]
 \psfrag{p}{$p_{\ell ab}({\rm GeV})\,\rightarrow$}
   \psfrag{s}{$\sigma(\rm m b)$}
  \centering
\includegraphics[scale=0.9]{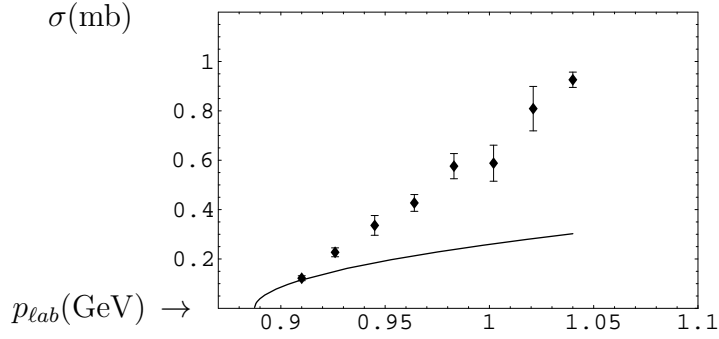}
\caption{The theoretical cross-section for the reaction $\pi^- p \to
  \Lambda^0 K^0$ with the $\pi^-p$ and $\Lambda^0K^0$ pairs in the
  S--wave state. The experimental points are taken from \cite{JJ71},
  $p_{\ell ab}$ is a momentum of the incident $\pi^-$--meson in the
  laboratory frame.}
\end{figure}
The theoretical value $A_{K^+\Lambda^0} = 1.22\,{\rm m\,b}/{\rm GeV}$
agrees well with the experimental data $A^{(\exp)}_{K^0\Lambda^0} =
(1.23 \pm 0.23)\,{\rm m\,b}/{\rm GeV}$ \cite{JJ71}. The cross section
for the reaction $\pi^-p \to K^0\Lambda^0$ Eq.(\ref{label7.8}) agrees
also well with the theoretical results obtained in \cite{WW97} within
$SU(3)$ chiral dynamics with coupled channels.

In the laboratory frame the threshold momentum of the $\pi^-$--meson
is equal to $p_{\rm th} = 0.887\,{\rm GeV}$. In the vicinity of
threshold the momentum $p_{K^0\Lambda^0}$ of the $K^0\Lambda^0$ pair
depends on the laboratory momentum $p_{\ell ab}$ of the $\pi^-$--meson
as follows
\begin{eqnarray}\label{label7.9}
  p_{K^0\Lambda^0} =  \sqrt{\frac{2 m^2_p m_{\Lambda^0} m_K\,( p^2_{\ell ab} - p^2_{\rm th})}{
(m_{\Lambda^0} + m_K)^2((m_{\Lambda^0} + m_K)^2 
      - m^2_p - m^2_{\pi})}} = 0.472\,\sqrt{p^2_{\ell ab} - p^2_{\rm th}}.
\end{eqnarray}
It is measured in ${\rm GeV}$. For the cross section for the reaction
$\pi^-p \to K^0\Lambda^0$ as a function of $p_{\ell ab}$ we get
\begin{eqnarray}\label{label7.10}
 \sigma^{\pi^-p \to K^0 \Lambda^0}(p_{\ell ab}) = (0.60 \pm 0.11)\,\sqrt{p^2_{\ell ab} - p^2_{\rm th}}
\end{eqnarray}
For $p_{\ell ab} = 0.910\,{\rm GeV}$ the theoretical cross section
$\sigma^{\pi^-p\to K^0\Lambda^0}(p_{\ell ab}) = 0.117\,{\rm mb}$ agrees
well with the experimental one $\sigma^{\pi^-p\to K^0\Lambda^0}_{ \exp}(p_{\ell ab})
= 0.122\pm 0.010\,{\rm mb}$ \cite{JJ71}. 

One can see that unlike the reactions $pN \to N\Lambda^0K^+$ the
contribution of the resonances $N(1535)$ and $N(1650)$ is very
important for the correct description of the reactions $\pi^0 p \to
\Lambda^0K^+$ and $\pi^-p \to \Lambda^0 K^0$ near threshold of the
final states \cite{WW97}. Indeed, the neglect of the resonance
contributions leads to the increase of the cross section for the
reaction $\pi^- p\to \Lambda^0K^+$ by a factor 3. This disagrees with
the experimental data \cite{JJ71}. The importance of the contributions
of the resonances $N(1535)$ and $N(1650)$ is caused by the substantial
cancellation of the non--resonance contributions described by the
first two terms in r.h.s. of Eq.(\ref{label7.5}).

In Fig.\,6 we show that the region of the applicability of our
theoretical cross section is restricted from above by the laboratory
momentum $p_{\ell ab} \le 0.910\,{\rm GeV}$. This corresponds to $0 \le
p_{K^0\Lambda^0} \le 0.095\,{\rm GeV}$. In terms of the energy excess
this region is defined by $0\le \varepsilon \le 13\,{\rm MeV}$, where
$\varepsilon = p^2_{K^0\Lambda^0}/2\mu$ and $\mu =
m_Km_{\Lambda^0}/(m_K + m_{\Lambda^0})$ is the reduced mass of the
$K^0\Lambda^0$ pair. It is obvious that the curve in Fig.6 is a branch
of hyperbola.

\section{Conclusion}
\setcounter{equation}{0}

Using chiral Lagrangians with linear realisation of chiral
$SU(3)\times SU(3)$ symmetry and non--derivative meson--baryon
couplings we have calculated the cross sections for the reactions $pN
\to N \Lambda^0 K^+$ near threshold of the final $N\Lambda^0K^+$
state. We have taken into account the contributions of the rescattering
in the initial $pN$ state and the final--state interactions in the
$N\Lambda^0$ channels. We have shown that the ratio of the cross
sections is equal to $R \simeq 3$. This agrees with an enhancement of
the strangeness production in the $pn$ interaction with respect to the
strangeness production in the $pp$ interaction observed by the
ANKE--Collaboration at COSY \cite{COSY}. We have explained such an
enhancement by the contribution of the $pn$ interaction in the
isospin--singlet state, which is much stronger than the $pn$
interaction in the isospin--triplet state.

We have shown that the contribution of the vector--meson exchanges is
at the level of a few percent. Hence, the vector\,--\,meson exchanges
cannot be fully responsible for the dynamics of strong low--energy
$pN$ interactions near threshold of the reaction $pN \to
N\Lambda^0K^+$. We have also shown that the contribution of the
resonances $N(1535)$ and $N(1650)$, treated as elementary particles,
to the matrix elements of the $pN \to N\Lambda^0K^+$ transitions does
not exceed $8\,\%$.  However, such an assertion should not be valid
within Wilkin's approach \cite{CW97,CW05}, where the main
contributions to the amplitudes of the reactions $pN \to
N\Lambda^0K^+$ come from the resonances $N(1535)$ and $N(1650)$,
described by the Breit--Wigner shapes.

In addition to vector--meson exchanges and resonances $N(1535)$ and
$N(1650)$ one can estimate the contributions of the scalar meson
$f_0(980)$ and the tensor meson $f_2(1270)$ \cite{PDG04}. Following
\cite{VC03} and \cite{BR70,TL04} one finds that the contribution of
the scalar meson $f_0(980)$ makes up about $2\,\%$ \cite{VC03},
whereas the contribution of the tensor meson $f_2(1270)$ is about
$5.7\,\%$ for the coupling constants $g^{(1)}_{f_2 NN} \simeq 2$ and
$g^{(2)}_{f_2 NN} = 0$ \cite{BR70,TL04}.  Since the theoretical
accuracy of our approach is not better than $15\,\%$, one can drop the
contributions of vector, tensor, exotic scalar mesons and resonances
$N(1535)$ and $N(1650)$ with respect to the contributions of ground
state baryon octet coupled to octets of pseudoscalar mesons and their
chiral scalar partners, which are taken in the infinite mass limit.

For the confirmation of our results, obtained for the strangeness
production near threshold of $pN$ interactions, we have calculated the
cross sections for the reactions $pp \to pp\pi^0$, $\pi^0p \to
\Lambda^0K^+$ and $\pi^-p \to \Lambda^0K^0$.  The theoretical results
agree well with the experimental data.  Unlike the strangeness
production in the $pN$ interactions the contribution of resonances
$N(1535)$ and $N(1650)$ to the amplitudes of the reactions $\pi N \to
\Lambda^0K$ has turned out to be very important. This is caused by the
substantial cancellation of the non--resonance contributions, defined
by interactions of ground state baryon octet with octets of
pseudoscalar and scalar mesons.

\section{Acknowledgement}

We are grateful to Vladimir Koptev for calling our attention to the
problem, investigated in this paper, and numerous and fruitful
discussions stimulating this work. 

One of the authors (A. Ivanov) is grateful to Colin Wilkin, Alexander
Sibirtsev, Alexander Kudryavtsev, Vadim Baru and all participants of
the seminar at Institut f\"ur Kernphysik at Forschungszentrum in
J\"ulich for fruitful discussions and constructive critical comments.
A. Ivanov thanks Prof.  Str\"oher for warm hospitality extended to him
during his stay at Institut f\"ur Kernphysik at Forschungszentrum in
J\"ulich.  The assistance of Yurii Valdau during the stay of A.
Ivanov at Institut f\"ur Kernphysik is greatly appreciated.

We are grateful to Igor Strakovsky for the discussions and comments on
the SAID analysis of the experimental data on $pp$ scattering
\cite{SAID1,SAID2}.

\section*{Appendix A: Cross sections for the reaction $pN \to
  N\Lambda^0K^+$ accounting for the final--state interaction}
\renewcommand{\theequation}{A-\arabic{equation}}
\setcounter{equation}{0}

The amplitude of the reaction $pN \to N\Lambda^0K^+$, where
$N\Lambda^0$ pair in the S--wave state, is related to the matrix
element of the $\mathbb{T}$--matrix as follows
\begin{eqnarray}\label{labelA.1}
  M(pN \to N\Lambda^0K^+) = \lim_{TV \to \infty}\frac{\langle N\Lambda^0K^+|\mathbb{T}|pN\rangle}{VT},
\end{eqnarray}
where $TV$ is a 4--dimensional volume defined by
$(2\pi)^4\delta^{(4)}(0) = TV$ and the $\mathbb{T}$--matrix obeys the
unitary condition \cite{SS61}
\begin{eqnarray}\label{labelA.2}
\mathbb{T} - \mathbb{T}^{\dagger} = i\,\mathbb{T}^{\dagger}\mathbb{T}.
\end{eqnarray}
The amplitude $M(pN \to N\Lambda^0K^+)$, defined by
Eq.(\ref{labelA.1}), describes the total amplitude of the reaction $pN
\to N\Lambda^0K^+$ accounting for the all kinds of interactions.

In order to separate the contributions of the final--state interaction
we propose to follow \cite{FSI1} (see also \cite{FSI} and \cite{AI1}).
First we define the amplitude of the reaction $pN \to N\Lambda^0K^+$,
which is calculated by the diagrams depicted in Fig.1--Fig.4:
\begin{eqnarray}\label{labelA.3}
  M_0(pN \to N\Lambda^0K^+) = 
\lim_{TV \to \infty}\frac{\langle N\Lambda^0K^+(0)|\mathbb{T}|pN\rangle}{VT},
\end{eqnarray}
where $|N\Lambda^0K^+(0)\rangle$ is the wave function of the final
state with free particles \cite{FSI1}. The account for the
final--state interaction in the $N\Lambda^0$ channel with the
$N\Lambda^0$ pair in the S--wave state can be given by means of the
wave function $|N\Lambda^0K^+(k_{N\Lambda})\rangle$ defined by
\cite{FSI1}
\begin{eqnarray}\label{labelA.4}
|N\Lambda^0K^+(k_{N\Lambda})\rangle = e^{\textstyle i\delta_{N\Lambda^0}(k_{N\Lambda^0})}
\frac{\sin \delta_{N\Lambda^0}(k_{N\Lambda^0})}{k_{N\Lambda^0}(- a_{N\Lambda^0})}|N\Lambda^0K^+(0)\rangle,
\end{eqnarray}
where $k_{N\Lambda^0}$ is a relative moment of the $N\Lambda^0$ pair,
$\delta_{N\Lambda^0}(k_{N\Lambda^0})$ and $a_{N\Lambda^0}$ are the
phase shift and the S--wave scattering length of $N\Lambda^0$
scattering in the S--wave state.  As has been shown in \cite{IV4} (see
also \cite{AI1}) such a wave function (\ref{labelA.4}) can be obtained
by summing up an infinite series of bubble--diagrams in the
$N\Lambda^0$ channel.

The wave function $|N\Lambda^0K^+(0)\rangle$ is normalised by \cite{IV4}
\begin{eqnarray}\label{labelA.5}
\hspace{-0.3in} && '\langle K^+(0)\Lambda^0 N|N\Lambda^0K^+(0)\rangle = 
(2\pi)^3 2E_{K^+}(\vec{p}_{K^+})\, 
\delta^{(3)}(\vec{p}_{K^+}\,' - \vec{p}_{K^+})\nonumber\\
\hspace{-0.3in}&&\times\,(2\pi)^3 2E_{\Lambda^0}(\vec{p}_{\Lambda^0}) 
\delta^{(3)}(\vec{p}_{\Lambda^0}\,' - \vec{p}_{\Lambda^0})\,\delta_{\sigma_{\Lambda^0}\,'\sigma_{\Lambda^0}}\,
(2\pi)^3 2E_N(\vec{p}_N) 
\delta^{(3)}(\vec{p}_N\,' - \vec{p}_N)\,\delta_{\sigma_N\,'\sigma_N},
\end{eqnarray}
where $\vec{p}_{K^+}, \vec{p}_X$ and $\vec{p}_{K^+}\,',\vec{p}_X\,'$
with $X =\Lambda^0, N$ are momenta of particles in the initial and
final states, $E_{K^+}(\vec{p}_{K^+})$ is the $K^+$--meson energy,
$E_X(\vec{p}_X)$ and $\sigma_X$ are the energy and polarisation of
the particle $X$.  Due to (\ref{labelA.5}) the wave function
$|N\Lambda^0K^+(k_{N\Lambda})\rangle$ is normalised as follows
\begin{eqnarray}\label{labelA.6}
\hspace{-0.3in} && '\langle K^+(k_{N\Lambda^0})\Lambda^0 N|N\Lambda^0K^+(k_{N\Lambda^0})\rangle = 
\frac{\sin^2 \delta_{N\Lambda^0}(k_{N\Lambda^0})}{k^2_{N\Lambda^0}a^2_{N\Lambda^0}}\,
(2\pi)^3 2E_{K^+}(\vec{p}_{K^+})\, 
\delta^{(3)}(\vec{p}_{K^+}\,' - \vec{p}_{K^+})\nonumber\\
\hspace{-0.3in}&&\times\,(2\pi)^3 2E_{\Lambda^0}(\vec{p}_{\Lambda^0}) 
\delta^{(3)}(\vec{p}_{\Lambda^0}\,' - \vec{p}_{\Lambda^0})\,\delta_{\sigma_{\Lambda^0}\,'\sigma_{\Lambda^0}}\,
(2\pi)^3 2E_N(\vec{p}_N) 
\delta^{(3)}(\vec{p}_N\,' - \vec{p}_N)\,\delta_{\sigma_N\,'\sigma_N} = P(k_{N\Lambda^0})\nonumber\\
\hspace{-0.3in}&&\times\,(2\pi)^3 2E_{\Lambda^0}(\vec{p}_{\Lambda^0}) 
\delta^{(3)}(\vec{p}_{\Lambda^0}\,' - \vec{p}_{\Lambda^0})\,\delta_{\sigma_{\Lambda^0}\,'\sigma_{\Lambda^0}}\,
(2\pi)^3 2E_N(\vec{p}_N) 
\delta^{(3)}(\vec{p}_N\,' - \vec{p}_N)\,\delta_{\sigma_N\,'\sigma_N}.
\end{eqnarray}
The  factor $P(k_{N\Lambda^0})$ can be rewritten in terms of the
cross sections for $N\Lambda^0$ scattering in the S--wave state
\begin{eqnarray}\label{labelA.7}
  P(k_{N\Lambda^0}) =  \frac{\sin^2 \delta_{N\Lambda^0}(k_{N\Lambda^0})}{k^2_{N\Lambda^0}a^2_{N\Lambda^0}}= 
  \frac{\sigma^{N\Lambda^0\to N\Lambda^0}(k_{N\Lambda^0})}{\sigma^{N\Lambda^0\to N\Lambda^0}(0)} \le 1,
\end{eqnarray}
where the cross sections $\sigma^{N\Lambda^0\to
  N\Lambda^0}(k_{N\Lambda^0})$ and $\sigma^{N\Lambda^0\to
  N\Lambda^0}(0)$ are equal to
\begin{eqnarray}\label{labelA.8}
 \sigma^{N\Lambda^0\to N\Lambda^0}(k_{N\Lambda^0}) =  4\pi\,
\frac{\sin^2 \delta_{N\Lambda^0}(k_{N\Lambda^0})}{k^2_{N\Lambda^0}}\quad,\quad
\sigma^{N\Lambda^0\to N\Lambda^0}(0) =  4\pi\,a^2_{N\Lambda^0}.
\end{eqnarray}
The factor $P(k_{N\Lambda^0})$ can be interpreted as a probability to
find a non--interacting system $N\Lambda^0 K^+$ with a relative
momentum $k_{N\Lambda^0}$ of the $N\Lambda^0$ pair \cite{SS61}. For
$k_{N\Lambda^0} = 0$, when the particles $N$ and $\Lambda^0$ are
separated by an infinite relative distance $r_{N\Lambda^0} \sim
1/k_{N\Lambda^0} \to \infty$ at $k_{N\Lambda^0} \to 0$, the probability
to find non--interacting particles in the state $|N\Lambda^0
K^+\rangle$ is equal to unity, i.e. $P(k_{N\Lambda^0} = 0) =1$. For
any finite $k_{N\Lambda^0}$ it is less than unity , i.e.
$P(k_{N\Lambda^0}) \le 1$ \cite{SS61}.

The amplitude of the reaction $pN \to N \Lambda^0 K^+$ accounting for
the final--state interaction in the $N\Lambda^0$ channel is defined by
\cite{FSI1}
\begin{eqnarray}\label{labelA.9}
  &&M(pN \to N\Lambda^0K^+) = \lim_{TV \to \infty}\frac{\langle N\Lambda^0K^+(k_{N\Lambda})|\mathbb{T}
    |pN\rangle}{VT} =\nonumber\\
  &&= \lim_{TV \to \infty}e^{\textstyle -\,i\delta_{N\Lambda^0}(k_{N\Lambda^0})}
  \frac{\sin \delta_{N\Lambda^0}(k_{N\Lambda^0})}{k_{N\Lambda^0}(- a_{N\Lambda^0})}\,
  \frac{\langle N\Lambda^0K^+(0)|\mathbb{T}|pN\rangle}{VT}=\nonumber\\
  &&=  e^{\textstyle -\,i\delta_{N\Lambda^0}(k_{N\Lambda^0})}
  \frac{\sin \delta_{N\Lambda^0}(k_{N\Lambda^0})}{k_{N\Lambda^0}(- a_{N\Lambda^0})}\,M_0(pN \to N\Lambda^0K^+).
\end{eqnarray}
The same result can be obtained by summing up an infinite series of
bubble diagrams in the $N\Lambda^0$ channel \cite{IV4} (see also
\cite{AI1}).  Hence the amplitude of the reaction $pN \to
N\Lambda^0K^+$ taking into account the final--state interaction in the
$N\Lambda^0$ channel is
\begin{eqnarray}\label{labelA.10}
 \hspace{-0.3in} M(pN \to N\Lambda^0K^+) = e^{\textstyle -\,i\delta_{N\Lambda^0}(k_{N\Lambda^0})}
  \frac{\sin \delta_{N\Lambda^0}(k_{N\Lambda^0})}{k_{N\Lambda^0}(- a_{N\Lambda^0})}\,M_0(pN \to N\Lambda^0K^+),
\end{eqnarray}
where the amplitude $M_0(pN \to N\Lambda^0K^+)$ is defined by the
Feynman diagrams in Fig.1--Fig.4.

At $k_{N\Lambda^0} = 0$ we get $M(pN \to N\Lambda^0K^+) = M_0(pN \to
N\Lambda^0K^+)$.  The reduction of the amplitude of the final--state
$N \Lambda^0$ interaction to unity at $k_{N\Lambda^0} = 0$ is clear
due to the short--range character of strong interactions. Indeed, at
$k_{N\Lambda^0} = 0$, corresponding to infinitely large relative
distance of the baryons in the $N\Lambda^0$ pair, the baryons do not
influence each other, since the wave functions of them do not overlap.

The contribution of the Coulomb interaction can be taken into account
according to the prescription \cite{FSI1} and Balewski {\it et al.}
\cite{FSI,JB98}.  This gives
\begin{eqnarray}\label{labelA.11}
 \hspace{-0.3in} M(pN \to N\Lambda^0K^+) = e^{\textstyle - i\delta_{N\Lambda^0}(k_{N\Lambda^0})}
  \frac{\sin \delta_{N\Lambda^0}(k_{N\Lambda^0}}{k_{N\Lambda^0}(- a_{N\Lambda^0})}\, 
C_0(k_{NK^+}) M_0(pN \to N\Lambda^0K^+).
\end{eqnarray}
Here $C_0(k_{NK^+})$ is the Gamow factor \cite{MN79,FSI1}
\begin{eqnarray}\label{labelA.12}
  C^2_0(k_{NK^+}) = \frac{2\pi\alpha Z_N\eta}{\displaystyle e^{\textstyle 2\pi\alpha Z_N\eta} - 1},
\end{eqnarray}
where $\alpha = 1/137.036$ is the fine--structure constant, $k_{NK^+}$
is a relative momentum of the $NK^+$ pair, $\eta = \mu_{NK^+}/k_{NK^+}
= 1/v_{NK^+}$ is the inverse relative velocity of the $N K^+$ pair,
$Z_N = 1$ for $N = p$ and $Z_N = 0$ for $N = n$. 

In the case of the Coulomb interaction the wave function
$|N\Lambda^0K^+\rangle$ has the normalisation (\ref{labelA.6}) with
the factor $P(k_{N\Lambda^0},k_{NK^+})$ equal to
\begin{eqnarray}\label{labelA.13}
  P(k_{N\Lambda^0},k_{NK^+}) =  C^2_0(k_{NK^+})\,
\frac{\sin^2 \delta_{N\Lambda^0}(k_{N\Lambda^0})}{k^2_{N\Lambda^0}a^2_{N\Lambda^0}} \le 1.
\end{eqnarray}
It defines the probability to find a non--interacting system
$N\Lambda^0K^+$ at relative momenta $k_{N\Lambda^0}$ and $k_{NK^+}$.

The cross section for the reaction $pN \to N\Lambda^0K^+$ with the
account for the final--state interaction is defined by
\begin{eqnarray}\label{labelA.14}
  \sigma^{pN \to N\Lambda^0K^+}(\varepsilon) = A_{pN}\,\varepsilon^2\,
  \Omega_{N\Lambda^0K^+}(\varepsilon),
\end{eqnarray}
where $A_{pN}$ is a constant, calculated at threshold momentum.  

The definition (\ref{labelA.14}) is correct in the vicinity of
threshold, where for the calculation of the amplitude $M_0(pN \to
N\Lambda^0K^+)$ of the reaction $pN \to N\Lambda^0K^+$ (or the
constant $A_{pN}$) one can neglect the dependence on the momenta of
the particle in the final state $N\Lambda^0K^+$. The function
$\Omega_{N\Lambda^0K^+}(\varepsilon)$ is determined by
\begin{eqnarray}\label{labelA.15}
  \Omega_{N\Lambda^0K^+}(\varepsilon) &=& \frac{2}{\pi^3}\int f_{\rm FSI}(\varepsilon,x)\,
\delta(1 - \vec{x}^{\,2} - \vec{y}^{\,2}\,)\,C^2_0(\vec{x},\vec{y}\,)d^3x d^3y,
\end{eqnarray}
where the function $f_{FSI}(\varepsilon,x)$ describes the final--state
interaction in the $N\Lambda^0$ channel \cite{FSI}.  It is normalised
to unity at $\varepsilon = 0$ and equal to \cite{FSI,FSIa}
\begin{eqnarray}\label{labelA.16}
  f_{\rm FSI}(\varepsilon,x) = \frac{1 }{\displaystyle 
    \Big(1 - \frac{\varepsilon m_{\Lambda^0}m_N}{m_{\Lambda^0} + m_N}\,a_{N\Lambda^0}r_{N\Lambda^0}x^2\Big)^2 + 
    \frac{2 \varepsilon m_{\Lambda^0}m_N}{m_{\Lambda^0} + m_N}\,a^2_{N\Lambda^0}x^2},
\end{eqnarray}
where $a_{N\Lambda^0} = -\,2\,{\rm fm}$ and $r_{N\Lambda^0} = 1\,{\rm
  fm}$ are the S--wave scattering length and effective range of
$N\Lambda^0$ scattering \cite{FSI}.

The Gamow factor depends on the variables $\vec{x}$ and $\vec{y}$ as
follows
\begin{eqnarray}\label{labelA.17}
  \eta = \sqrt{\frac{m_p (m_{\Lambda^0} + m_p)}{2\varepsilon m_{\Lambda^0}}}\,
  \frac{1}{\displaystyle \Big| \vec{x} - 
    \sqrt{\frac{m_p (m_{\Lambda^0} + m_p + m_K)}{m_{\Lambda^0} m_K}}\,\vec{y}\,\Big|} = 
\frac{1}{v}\,\sqrt{\frac{m_p (m_{\Lambda^0} + m_p)}{2\varepsilon m_{\Lambda^0}}}.
\end{eqnarray}
The function $\Omega_{N \Lambda^0K^+}(\varepsilon)$ is equal to to
unity for $\alpha = 0$ and $a_{N\Lambda^0} = r_{N\Lambda^0} = 0$.  For
the reactions $pp \to p \Lambda^0 K^+$ and $pn \to n\Lambda^0 K^+$ we
get
\begin{eqnarray}\label{labelA.18}
  \hspace{-0.3in}\Omega_{p\Lambda^0K^+}(\varepsilon) =  \sqrt{\frac{m_K(m_{\Lambda^0} + m_p)}{2 
      \varepsilon (m_{\Lambda^0} + m_p + m_K)}}
  \frac{8}{\pi}\int^1_0  \int^{v_+(x)}_{v_-(x)}\!\!\!\frac{f_{\rm FSI}(\varepsilon,x)2\pi \alpha dv x dx}{\displaystyle 
    \exp\Big\{\frac{2\pi \alpha}{v}
    \sqrt{\frac{m_p(m_{ \Lambda^0} + m_p)}{2\varepsilon m_{\Lambda^0}}}\Big\} - 1},
\end{eqnarray}
where we have denoted
\begin{eqnarray}\label{labelA.19}
  v_+ &=& x + \sqrt{\frac{m_p (m_{\Lambda^0} + m_p + m_K)}{m_{\Lambda^0} m_K}}\,\sqrt{1 - x^2},\nonumber\\
  v_- &=& \Big|x -  \sqrt{\frac{m_p (m_{\Lambda^0} + m_p + m_K)}{m_{\Lambda^0} m_K}}\,\sqrt{1 - x^2}\Big| 
\end{eqnarray}
and
\begin{eqnarray}\label{labelA.20}
  \Omega_{n\Lambda^0K^+}(\varepsilon) =
  \frac{16}{\pi}\int^1_0
  f_{\rm FSI}(\varepsilon,x)\,x^2 \sqrt{1 - x^2}\,dx.
\end{eqnarray}
For $\varepsilon = 6.68\,{\rm MeV}$ the contribution of the Coulomb
repulsion in the $K^+p$ pair to function
$\Omega_{p\Lambda^0K^+}(\varepsilon)$ makes up about $15\,\%$. At the
neglect of the contribution of the Coulomb repulsion
$\Omega_{p\Lambda^0K^+}(\varepsilon) =
\Omega_{n\Lambda^0K^+}(\varepsilon) = 0.67$ at $\varepsilon =
6.68\,{\rm MeV}$.

In the reaction $pp \to pp\pi^0$ the account for the final--state
interaction in the $pp$ channel has been carried out within the scheme
expounded above. As we have shown the theoretical cross sections for
the reactions $pp \to p \Lambda^0 K^+$ and $pp \to pp \pi^0$ agree
well with the experimental data.


\begin{thebibliography}{9}
\bibitem{COSY}  
M. B\"uscher {\it et al.},
Eur. Phys. J. A {\bf 22}, 301 (2004).
\bibitem{TE88} 
T. E. O. Ericson and W. Weise, 
in {\it PIONS AND NUCLEI}, Clarendon Press, Oxford, 1988.
\bibitem{EC94}
E. Chiavassa {\it et al.},
Phys. Lett. B {\bf 337}, 192 (1994).
\bibitem{KT99}
K. Tsushima {\it et al.},
Phys. Rev. C {\bf 59}, 369 (1999).
\bibitem{VK05}
V. Koptev (private communication), 2005.
\bibitem{CW97}
G. F\"aldt and C. Wilkin,
Eur. Phys. J. A {\bf 24}, 431 (2005), nucl--th/0411019;
 Z. Phys. A {\bf 357}, 241 (1997).
\bibitem{PDG04} 
S. Eidelman {\it et al.} (The Particle Data Group),
Phys. Lett. B {\bf 595}, 1 (2004).
\bibitem{CW05}
C. Wilkin (private communication), J\"ulich 2005.
\bibitem{AS05}
A. Sibirtsev (private communication), J\"ulich 2005.
\bibitem{AI01} 
A. Ya. Berdnikov {\it et al.}, Eur. Phys. J. A {\bf 9}, 425 (2000);
Eur. Phys. J. A {\bf 12}, 341 (2001).
\bibitem{HP73}
V. De Alfaro, S. Fubini, G. Furlan, and C. Rossetti,
in {\it Currents in Hadron Physics},
North--Holland Publishing Co., Amsterdam $\,\bullet\,$ London,
American Elsevier Publishing Co., Inc.,
New York, 1973.
\bibitem{SW67}
S. Weinberg,
Phys. Rev.  Lett. {\bf 18}, 188 (1967).
\bibitem{BL69}
W. Bardeen and B. W. Lee,
Phys. Rev. {\bf 177}, 2389 (1969).
\bibitem{GG69}
S. Gasiorowicz and D. A. Geffen, 
Rev. Mod. Phys. {\bf 41}, 531 (1969).
\bibitem{NK99}
N. Kaiser,
Eur. Phys. J. A {\bf 5}, 105 (1999).
\bibitem{KN99}
C. Hanhart and K. Nakayama,
Phys. Lett. B {\bf 454}, 176 (1999).
\bibitem{FSI}
J. T. Balewski {\it et al.},
Eur. Phys. J. A {\bf 2}, 99 (1998) and references therein.
\bibitem{FSIa}
(see \cite{FSI} Eq.(4)).
\bibitem{IV4}
A. N. Ivanov {\it et al.},
Eur. Phys. J. A {\bf 23}, 79 (2005); nucl--th/0406053.
\bibitem{JB98}
J. T. Balewski {\it et al.},
Phys. Lett. B {\bf 420}, 211 (1998). 
\bibitem{MR96}
C.--H. Lee, D.--P. Min, and M. Rho,
Phys. Lett. B {\bf 326}, 14 (1994);
C.--H. Lee, G. E. Brown, and M. Rho,
Phys. Lett. B {\bf 335}, 266 (1994);
C.--H. Lee, G. E. Brown, D.--P. Min, and M. Rho,
Nucl. Phys. A {\bf 585}, 401 (1995);
C.--H. Lee, D.--P. Min, and M. Rho,
Nucl. Phys. A {\bf 602}, 334 (1996).
\bibitem{AI99}
A. N. Ivanov,
 M. Nagy, and N. I. Troitskaya,
Phys. Rev.  {\bf C59}, 451 (1999) and references therein.
\bibitem{TE05}
T. E. O. Ericson and A. N. Ivanov,
Phys. Lett. B {\bf 634}, 39 (2006).
\bibitem{IV3}
A. N. Ivanov {\it et al.},
Eur. Phys. J. A {\bf 21}, 11 (2004); nucl--th/0310081.
\bibitem{IV5}
 A. N. Ivanov {\it et al.},
Phys. Rev. A {\bf 71}, 052508  (2005).
\bibitem{IV8} 
A. N. Ivanov {\it et al.}, 
Eur. Phys. J. A {\bf 25}, 329 (2005), nucl\,--\,th/0505078.
\bibitem{IV9}
A. N. Ivanov {\it et al.},  
{\it  Recent Theoretical Studies on Hadronic Atoms},
Invited talk at International Conference on Exotic Atoms -- EXA05 
at Stefan Meyer Institute of Austrian Academie of Sciences, Vienna, 
Austria, 21-25 February 2005, nucl--th/0505022.
\bibitem{JJ71}
J. J. Jones {\it et al.},
Phys. Rev. Lett. {\bf 26}, 860 (1971).
\bibitem{WW97}
N. Kaiser, T. Waas, and W. Weise,
Nucl. Phys. A {\bf 612}, 297 (1997).
\bibitem{MN79}
M. M. Nagels {\it et al.} 
{\it Compilation of coupling constants and low--energy parameters},
Nucl. Phys. B {\bf 147}, 189 (1979);
O. Dumbrajs {\it et al.}, 
{\it Compilation of coupling constants and low--energy parameters},
Nucl. Phys. B {\bf 216}, 277 (1983).
\bibitem{PSI2}
H.--Ch. Schr\"oder {\it et al.},
Eur. Phys. J. C {\bf 21}, 473 (2001).
\bibitem{SAID1}
R. A. Arndt, I. I. Strakovsky, and R. L. Workman,
Phys. Rev. C {\bf 62}, 034005 (2000), nucl--th/0004039;
http://gwdac.phys.gwu.edu
\bibitem{SAID2}
I. I. Strakovsky (private communication), 2005.
\bibitem{PDG04a}
(see \cite{PDG04} p.872)
\bibitem{PDG04b}
(see \cite{PDG04} p.158)
\bibitem{HM1}
H. O. Meyer {\it et al.},
Nucl. Phys. A {\bf 539}, 633 (1992).
\bibitem{HM2}
H. Machner and J. Haidenbauer,
J. Phys. G: Nucl. Part. Phys. {\bf 25}, R231 (1999).
\bibitem{NK01}
V. Bernard, N. Kaiser, and Ulf.--G. Mei\ss ner,
Eur. Phys. A {\bf 4}, 259 (1999).
\bibitem{NK02}
C. Hanhart and N. Kaiser,
Phys. Rev. C {\bf 66}, 054005 (2002) and references therein.
\bibitem{AI1}
A. N. Ivanov, N. I. Troitskaya, H. Oberhummer, and M. Faber,
Eur. Phys. J.  A {\bf 8}, 223  (2000).
\bibitem{KM58}
K. B. Mather and P. Swan,
in {\it Nuclear scattering}, Cambridge at the University Press, 1958, p.214.
\bibitem{VC03}
V. Cirigliano, G. Ecker, H. Neufeld, A. Pich,
 JHEP {\bf 0306}, 012 (2003).
\bibitem{BR70}
B. Renner,
Phys. Lett. B {\bf 33}, 599 (1970).
\bibitem{TL04}
Y. Oh and T.--S. H. Lee,
Phys. Rev. C {\bf 69}, 025201 (2004).
\bibitem{SS61}
S. S. Schweber,
in {\it An introduction to relativistic field theory},
Row, Peterson and Co$\,\bullet\,$ Evanston, Ill.,
Elmsford, New York, 1961.
\bibitem{FSI1}
(see also \cite{HM2} p.241).
\end{thebibliography}
\end{document}